\documentclass[10pt]{article}

\usepackage{amsmath,latexsym,color, subfigure, url,setspace, multirow}
\usepackage{amssymb}

\usepackage{graphicx}

\usepackage{cite}

\usepackage[labelfont=bf,labelsep=period,justification=raggedright]{caption}

\makeatletter
\renewcommand{\@biblabel}[1]{\quad#1.}
\makeatother

\date{}

\newcommand{\bx}{\mathbf{x}}
\newcommand{\by}{\mathbf{y}}
\newcommand{\bz}{\mathbf{z}}
\newcommand{\bu}{\mathbf{u}}

\newcommand{\bK}{\mathbf{K}}
\newcommand{\bX}{\mathbf{X}}

\newcommand{\bH}{\mathbf{H}}
\newcommand{\bP}{\mathbf{P}}

\newcommand{\bM}{\mathbf{M}}
\newcommand{\bU}{\mathbf{U}}
\newcommand{\bD}{\mathbf{D}}
\newcommand{\bI}{\mathbf{I}}
\newcommand{\bepsilon}{\boldsymbol\epsilon}
\newcommand{\bbeta}{\boldsymbol\beta}
\newcommand{\tbbeta}{{\tilde{\boldsymbol\beta}}}
\newcommand{\tbeta}{{\tilde{\beta}}}
\newcommand{\bgamma}{\boldsymbol\gamma}
\newcommand{\bdelta}{\boldsymbol\delta}
\newcommand{\balpha}{\boldsymbol\alpha}
\newcommand{\bOmega}{\boldsymbol\Omega}
\begin{document}

% Title must be 150 characters or less
\begin{flushleft}
{\Large
\textbf{Polygenic Modeling with Bayesian Sparse Linear Mixed Models}
}
% Insert Author names, affiliations and corresponding author email.
\\
Xiang Zhou$^{1,\ast}$, 
Peter Carbonetto$^{1}$, 
Matthew Stephens$^{1,2,\ast}$
\\
\bf{1} Dept. of Human Genetics, University of Chicago, Chicago, IL, USA
\\
\bf{2} Dept. of Statistics, University of Chicago, Chicago, IL, USA
\\
$\ast$ E-mail: xz7@uchicago.edu; mstephens@uchicago.edu
\end{flushleft}

% Please keep the abstract between 250 and 300 words
\section*{Abstract}
Both linear mixed models (LMMs) and sparse regression models are widely used in genetics applications, including, recently, 
polygenic modeling in genome-wide association studies. These two approaches make very different assumptions, so are expected to perform well
in different situations. However, in practice, for a given data set one typically does not know which assumptions will be more accurate.
Motivated by this, we consider a hybrid of the two, which we refer to as a ``Bayesian sparse linear mixed model" (BSLMM) that includes
both these models as special cases. We address several key computational and statistical issues that arise when applying BSLMM, including appropriate prior specification for the hyper-parameters, and a novel Markov chain Monte Carlo algorithm for posterior inference. We apply BSLMM and compare it with other methods for two polygenic modeling applications: estimating the proportion of variance in phenotypes explained (PVE) by available genotypes, and phenotype (or breeding value) prediction. For PVE estimation, we demonstrate that BSLMM combines the advantages of both standard LMMs and sparse regression modeling. For phenotype prediction it considerably outperforms either of the other two methods, as well as
several other large-scale regression methods previously suggested for this problem. Software implementing our method is freely available from \url{http://stephenslab.uchicago.edu/software.html}.
% Please keep the Author Summary between 150 and 200 words
% Use first person. PLoS ONE authors please skip this step. 
% Author Summary not valid for PLoS ONE submissions.   
\section*{Author Summary}
The goal of polygenic modeling is to better understand the relationship between genetic variation
and variation in observed characteristics, including variation in quantitative traits (e.g. cholesterol level in humans, milk production in cattle) and disease susceptibility. Improvements in polygenic modeling will help 
improve our understanding to this relationship, and could ultimately lead to, for example, changes in 
clinical practice in humans, or better breeding/mating strategies in agricultural programs.
Polygenic models present important challenges, both at the modeling/statistical level (what modeling assumptions produce the best results), and at the computational level (how should these models
be effectively fit to data). We develop novel approaches to help tackle both these challenges, and demonstrate the gains in accuracy that result on both simulated and real data examples.

\clearpage
\newpage
\section*{Introduction}
Both linear mixed models (LMMs) and sparse regression models are widely used in genetics applications.  
For example, LMMs are often used to control for population stratification, individual relatedness, or unmeasured 
confounding factors when performing association tests in genetic association studies \cite{Abney:2002, Yu:2006, Aulchenko:2007, Kang:2008, Kang:2010, Zhang:2010,Price:2010, Lippert:2011, Zhou:2012} and gene expression studies \cite{Kang:2008b, Listgarten:2010, Fusi:2012}.
They have also been used in genetic association studies to jointly analyze groups of SNPs \cite{Malo:2008, Chen:2010}. 
Similarly, sparse regression models have been used in genome-wide association analyses 
\cite{Yi:2008, Hoggart:2008, Wu:2009, Logsdon:2010, Guan:2011, Carbonetto:2012} and 
in expression QTL analysis \cite{LeeSI:2009}. Further, both LMMs and sparse regression models have been applied to, and garnered renewed interest in, polygenic modeling in association studies. Here, by polygenic modeling we mean any attempt to relate phenotypic variation to many genetic variants simultaneously (in contrast to single-SNP tests of association).
The particular polygenic modeling problems that we focus on here are
estimating ``chip heritability", being the proportion of variance in phenotypes explained (PVE) by available genotypes \cite{Yang:2010, Golan:2011, Lee:2011, Guan:2011}, and predicting phenotypes based on genotypes \cite{Henderson:1984, Whittaker:2000, Meuwissen:2001, Makowsky:2011, Ober:2012}. 

Despite the considerable overlap in their applications, in the context of polygenic modeling, LMMs and sparse regression models are based on almost diametrically opposed assumptions. Precisely, applications of LMMs to polygenic modeling (e.g.~\cite{Yang:2010}) effectively assume that every genetic variant affects the phenotype, with
effect sizes normally distributed, whereas sparse regression models, such as Bayesian variable selection regression models (BVSR) \cite{Logsdon:2010, Guan:2011}, assume that a relatively small proportion of all variants
affect the phenotype. The relative performance of these two models for polygenic modeling applications
would therefore be expected to vary depending on the true underlying genetic architecture of the phenotype. However, in practice, one does not know the true genetic architecture, so it is unclear which
of the two models to prefer. Motivated by this observation, we consider a hybrid of these two models, which
we refer to as the ``Bayesian sparse linear mixed model", or BSLMM. This hybrid includes both the LMM and a sparse regression model, BVSR, 
as special cases, and is to some extent capable of learning the genetic architecture from the data, yielding
good performance across a wide range of scenarios. By being ``adaptive" to the data in this way, our approach 
obviates the need to choose one model over the other, and attempts to combine the benefits of both.

The idea of a hybrid between LMM and sparse regression models is, in itself, not new. Indeed, models like these have been used in breeding value prediction to assist genomic selection in animal and plant breeding programs \cite{Piepho:2009, Goddard:2009, Verbyla:2009, Verbyla:2010, Habier:2011, Erbe:2012}, gene selection in expression analysis while controlling for batch effects \cite{Baragatti:2011}, phenotype prediction of complex traits in model organisms and dairy cattle  \cite{Lee:2008, Legarra:2008, Campos:2009, Hayes:2010}, and more recently, mapping complex traits by jointly modeling all SNPs in structured populations \cite{Segura:2012}. Compared with these previous papers, our work makes two key contributions. 
First, we consider in detail the specification of appropriate prior distributions for the hyper-parameters of the model.
We particularly emphasize the benefits of estimating the hyper-parameters from the data, rather than fixing them to pre-specified values
to achieve the adaptive behavior mentioned above. Second, we provide a novel computational algorithm that exploits a recently described linear algebra trick for LMMs \cite{Lippert:2011, Zhou:2012}. The resulting algorithm
avoids {\it ad hoc} approximations that are sometimes made when fitting these types of model (e.g. \cite{Lee:2008, Segura:2012} ), and yields
reliable results for data sets containing thousands of individuals and hundreds of thousands of markers. (Most previous applications of this kind of model involved much smaller data sets.) 

Since BSLMM is a hybrid of two widely used models, it naturally has
 a wide range of potential uses. Here we focus on its application to polygenic modeling for genome-wide association studies, 
 specifically two applications of particular interest and importance:  PVE estimation (or ``chip heritability" estimation) and phenotype prediction.
Estimating the PVE from large-scale genotyped marker sets (e.g. all the SNPs on a genome-wide genotyping chip) has the potential to shed light on sources of ``missing heritability" \cite{Eichler:2010} and the underlying genetic architecture of common diseases \cite{Yang:2010, Lee:2011, Golan:2011, Guan:2011, Stahl:2012}. And accurate prediction of complex phenotypes from genotypes could ultimately impact many areas of genetics, including applications in animal breeding, medicine and forensics \cite{Meuwissen:2001, Lee:2008, Legarra:2008, Campos:2009, Hayes:2010, Makowsky:2011, Ober:2012}. Through simulations and applications to real data, we show that BSLMM successfully combines the advantages of both LMMs and sparse regression, is robust to a variety of settings in PVE estimation, and outperforms both models, and several related models, in phenotype prediction. 

\clearpage
\newpage
\section*{Methods}

\subsection*{Background and Motivation}
We begin by considering a simple linear model relating phenotypes $\by$
to genotypes $\bX$:
\begin{align} \label{eqn:simple-linear}
\by&=\mathbf 1_n\mu+\bX \bbeta+\bepsilon,  \\
\bepsilon&\sim \mbox{MVN}_n(0, \bI_n \tau^{-1}).
\end{align}

Here $\by$ is an $n$-vector of phenotypes measured on $n$ individuals,
$\bX$ is an $n \times p$ matrix of genotypes measured on the same individuals
at $p$ genetic markers, $\bbeta$ is a $p$-vector of (unknown) genetic marker effects, $\mathbf 1_n$ is an $n$-vector of 1s, $\mu$ is a scalar representing the phenotype mean, and $\bepsilon$ is an $n$-vector of error terms that have variance $\tau^{-1}$. $\mbox{MVN}_n$ denotes the $n$-dimensional multivariate normal distribution. Note that there are many ways in which measured genotypes can be encoded in the matrix $\bX$. We assume throughout this paper that the genotypes are coded as 0, 1 or 2 copies of a reference allele at each marker, and that the columns of $\bX$ are centered but not standardized; see Text S1 Detailed Methods.

A key issue is that, in typical current datasets (e.g. GWAS), the number of markers $p$ is much larger than the number of individuals $n$.
As a result, parameters of interest (e.g. $\bbeta$ or PVE) cannot be estimated accurately without making some kind of modeling assumptions. Indeed, many existing approaches to polygenic modeling can be derived
from (\ref{eqn:simple-linear}) by making specific assumptions
about the genetic effects $\bbeta$. For example, the LMM approach
from \cite{Yang:2010}, which has recently become commonly used for PVE
estimation (e.g. \cite{Yang:2011, Lee:2011, Deary:2012, Lee:2012}), is equivalent to the assumption that
effect sizes are normally distributed, such that
\begin{equation} \label{eqn:normal}
\beta_i \sim \mbox{N}(0, \sigma_b^2/(p\tau)).
\end{equation}
[Specifically, exact equivalence holds when the relatedness matrix in the LMM is computed from the genotypes as $\bK=\frac{1}{p}\bX\bX^T$ (e.g. \cite{Campos:2010}). \cite{Yang:2010} use a matrix in this form, with $\bX$ centered and standardized, and with a slight modification of the diagonal elements.]
For brevity, in this paper we refer to the regression model that results from this assumption as the LMM (note that this is relatively restrictive compared with the usual definition);
it is also commonly referred to as ``ridge regression" in statistics \cite{Hoerl:1970}. The estimated combined effects ($\bX\bbeta$), or equivalently, the estimated random effects, obtained from this model are commonly referred to as Best Linear Unbiased Predictors (BLUP) \cite{Robinson:1991}.

An alternative assumption, which has also been widely used in polygenic modeling applications \cite{Logsdon:2010, Guan:2011, Habier:2011}, and more generally in statistics for sparse high-dimensional regression with large numbers of covariates  \cite{Clyde:1996, Chipman:2001}, is that the effects come from a mixture of a normal distribution and
a point mass at 0, also known as a point-normal distribution:
\begin{equation} \label{eqn:point-normal}
\beta_i \sim \pi \mbox{N}(0, \sigma_a^2/(p\tau))+(1-\pi)\delta_0,
\end{equation}
where $\pi$ is the proportion of non-zero $\bbeta$ and $\delta_0$ denotes a point mass at zero. [This definition of $\pi$ follows the convention from statistics \cite{Clyde:1996, Chipman:2001, Guan:2011}, which is opposite to the convention in animal breeding \cite{Meuwissen:2001, Verbyla:2009, Verbyla:2010, Hayes:2010, Habier:2011}.] We refer to the resulting regression model as Bayesian Variable Selection Regression (BVSR),
because it is commonly used to select the relevant variables (i.e.~those with non-zero effect) for phenotype prediction.  Although (\ref{eqn:point-normal}) formally includes 
 (\ref{eqn:normal}) as a special case when $\pi=1$, in practice (\ref{eqn:point-normal}) is often used together with an
assumption that only a small proportion of the variables are likely to be relevant for phenotype prediction, say by specifying a prior distribution for $\pi$ that puts appreciable mass on
small values (e.g.~\cite{Guan:2011}). In this case, BVSR and LMM can be viewed as making almost diametrically opposed
assumptions: the LMM assumes every variant has an effect, whereas BVSR assumes that
a very small proportion of variants have an effect. (In practice, the estimate of $\sigma_b$ under LMM is often smaller than the estimate of $\sigma_a$ under BVSR, so we can interpret the LMM as assuming a large number of small effects, and BVSR as assuming a small number of larger effects.)

A more general assumption, which includes both the above as special cases, is that the
effects come from a mixture of two normal distributions:
\begin{equation} \label{eqn:normal-mix}
\beta_i \sim \pi \mbox{N}(0, (\sigma_a^2+\sigma_b^2)/(p\tau))+(1-\pi)\mbox{N}(0, \sigma_b^2/(p\tau)),
\end{equation}
Setting $\pi=0$ yields the LMM (\ref{eqn:normal}), and $\sigma_b=0$ yields BVSR (\ref{eqn:point-normal}).
we can interpret this model as assuming that all variants have at least a small effect, which are normally distributed with variance $\sigma_b^2/(p\tau)$, and some proportion ($\pi$) of variants have an additional effect, normally distributed with variance $\sigma_a^2/(p\tau)$. 
The earliest use of a mixture of two normal distributions for the regression
coefficients that we are aware of is \cite{George:1993}, although in that paper
various hyper-parameters were fixed, and so it did not include LMM and BVSR as special cases. 

Of the three assumptions on the effect size distributions above, the
last (\ref{eqn:normal-mix}) is clearly the most flexible since it includes the others as special cases. Obviously other assumptions are possible, 
some still more flexible than the mixture of two normals: for example, a mixture of three or more normals. Indeed, many other assumptions have been proposed, including
variants in which a normal distribution is replaced by a $t$ distribution. 
These variants include the ``Bayesian alphabet models" -- so-called simply because they
have been given names such as BayesA, BayesB, BayesC, etc. -- that have been 
proposed for polygenic modeling, particularly breeding value prediction in genomic selection studies. 
Table \ref{tab:balpha} summarizes these models, and some other effect size distributions
that have been proposed, together with relevant references (see also \cite{Campos:2012} and the references there in). Among these, the models most closely related to ours are BayesC$\pi$ \cite{Habier:2011} and BayesR \cite{Erbe:2012}. Specifically, BayesC$\pi$ without a random effect is BVSR, and with a random effect is BSLMM (which we define below). BayesR models effect sizes using a mixture of three normal components plus a point mass at zero, although the relative variance for each normal distribution is fixed.

Given the wide range of assumptions for effect size distributions that have been proposed, it is natural to wonder which are the most appropriate for general use. However, answering this question is
complicated by the fact that even given the effect size distribution there are a number of different ways
that these models can be implemented in practice, both in terms of statistical issues, such as
treatment of the hyper-parameters, and in terms of computational and algorithmic issues.
Both these types of issues can greatly affect practical performance. For example, many approaches fix the hyper-parameters to specific values
\cite{Meuwissen:2001, Verbyla:2009, Verbyla:2010, Hayes:2010} which makes them less flexible \cite{Gianola:2009, Habier:2011}. Thus, in this paper we focus on a particular effect size distribution (\ref{eqn:normal-mix}), which while not the most flexible among all that could be considered, 
is certainly more flexible than the one that has been most used in practice for estimating PVE (LMM),
and admits certain computational methods that could not be applied in all cases.
We consider in detail how to apply this model in practice, and the resulting advantages over LMM and BVSR (although we also compare with some
other existing approaches). 
A key contribution of this paper is to provide
new approaches to address two important practical issues: the statistical issue of how to deal with the unknown hyper-parameters $(\pi, \sigma_a, \sigma_b)$, and the computational issue of how to fit the model. Notably, with the computational tools we use here, fitting the model (\ref{eqn:normal-mix}) becomes, for a typical data set, less computationally intensive than fitting BVSR, as well as
providing gains in performance.

With this background, we now turn to detailed description of the model, its prior specification and its computation algorithm.

\subsection*{The Bayesian Sparse Linear Mixed Model}

In this paper we focus on the simple linear model (\ref{eqn:simple-linear}) with mixture prior (\ref{eqn:normal-mix}) on the effects. However, the computational and statistical methods we use here also apply to a more general model,  which we
refer to as the Bayesian Sparse Linear Mixed Model (BSLMM), and which includes the model (\ref{eqn:simple-linear}) with (\ref{eqn:normal-mix}) as a special case.

The BSLMM consists of a standard linear mixed model, with one random effect term, and with sparsity inducing priors on the regression coefficients:
\begin{align}
\label{eqn_bslmm}
\by&=\mathbf 1_n \mu+\bX\tbbeta+\bu+\bepsilon,   \\  \label{eqn_uterm}
\bu&\sim \mbox{MVN}_n(0, \sigma_b^2 \tau^{-1} \bK), \\
\bepsilon &\sim \mbox{MVN}_n(0, \tau^{-1} \bI_n),\\
\tbeta_i &\sim \pi \mbox{N}(0, \sigma_a^2\tau^{-1})+(1-\pi)\delta_0,
\end{align}
where $\bu$ is an $n$-vector of random effects with known $n \times n$ covariance matrix $\bK$. In referring to $\bu$ as the ``random effects"
we are following standard terminology from LMMs. Standard terminology 
also refers to the coefficients
$\tbbeta$ as ``fixed effects", but this phrase has a range of potential meanings \cite{Gelman:2005}
and so we avoid it here. Instead we use the term ``sparse effects" for these parameters to emphasize the sparsity-inducing prior.

It is straightforward to show that when $\bK=\frac{1}{p}\bX\bX^T$, BSLMM is equivalent to
the simple linear model (\ref{eqn:simple-linear}) with mixture prior (\ref{eqn:normal-mix}) on the effects. However, our discussion of prior specification, computational algorithms, and software, all apply for any $\bK$.

%Here, we use the terminology random effect to refer to coefficients that vary for different groups and the terminology fixed effect to refer to coefficients that are identical for all groups, following conventional definition \cite{Gelman:2005}. Specifically, we term $\bu$ as random effects because we view each individual as a random sample from the population. As a result, the corresponding value $u_i$ for one individual is different from that for another. On the other hand, we term $\tbbeta$ as fixed effects because each genotype (element in $\bX$) can only exhibit one of a few possible values (i.e. 0/1/2) and a same genotype is repeatedly observed for many individuals. Consequently, $\tbbeta$ remain the same for every individual from the population. We caution that the difference between the two terms can be subtle, and that other and inconsistent definitions of the two are prevalent throughout literature (see a relevant discussion in \cite{Gelman:2005}). Therefore, we note that we only use these two terms for terminology purpose in the present study.} 

When we say that (\ref{eqn_bslmm}) is equivalent to (\ref{eqn:simple-linear}) with (\ref{eqn:normal-mix}), this equivalence refers to the implied probability model for $\by$ given $\bX$
and the hyper-parameters $\mu, \tau, \pi, \sigma_a,\sigma_b$. However, $\tbbeta$ and $\bbeta$ are not equivalent (explaining our use of two different symbols):
in (\ref{eqn_bslmm}) the random effect $\bu$ captures the combined small effects of all markers,
whereas in (\ref{eqn:simple-linear}) these small effects are included in $\bbeta$. Since both 
our applications focus on the relationship between $\by$ and $\bX$, and not on interpreting
estimates of $\bbeta$ or $\tbbeta$, we do not concern ourselves any further with this issue, although it
may need consideration in applications where individual estimated genetic effects are of more direct interest (e.g.~genetic association mapping).
A related issue is the interpretation of the random effect $\bu$ in BSLMM: 
from the way we have presented the material $\bu$ is most naturally
interpreted as representing a polygenic component, specifically the combined effect of a large number of small effects across all measured markers. However, if there are environmental
factors that influence the phenotype and are correlated with genotype (e.g.~due to population structure), 
then these would certainly affect estimates of $\bu$, and consequently also affect estimates of other quantities, including the PVE.  
In addition, phenotype predictions from BSLMM will include a component due to
unmeasured environmental factors that are correlated with measured genotypes. 
These issues are, of course, not unique to BSLMM -- indeed, they apply equally to the LMM;
see \cite{Browning:2011} and the response from \cite{Goddard:2011} for relevant discussion.

Finally, given the observation that a mixture of two normals is more flexible than a point-normal, it might seem natural to consider modifying (\ref{eqn_bslmm}) by making the assumption that $\tbbeta$ comes from a mixture of two normal distributions rather than a point-normal. However, if
$\bK=\frac{1}{p}\bX\bX^T$ then this modification is simply equivalent to changing
 the values of $\sigma_a,\sigma_b$.

\subsection*{Prior Specification}    

The BSLMM involves (hyper-)parameters, $\mu,\tau,\pi, \sigma_a,$ and $\sigma_b$.
Before
considering prior specification for these parameters, we summarize their interpretations as follows: 
\begin{itemize}
\item $\mu$ and $\tau^{-1}$ control the phenotype mean and residual variance.
\item $\pi$ controls the proportion of $\tbbeta$ values in (\ref{eqn_bslmm}) that are non-zero. 
\item $\sigma_a$ controls the expected magnitude of the (non-zero) $\tbbeta$.
\item $\sigma_b$ controls the expected magnitude of the random effects $\bu$.
\end{itemize}
Appropriate values for these parameters will clearly vary for different data sets, so it seems desirable to estimate them from the data. Here we accomplish this in a Bayesian framework by specifying prior distributions for the parameters, and using Markov chain Monte Carlo (MCMC) to obtain approximate samples from their posterior distribution given the observed data.
Although one could imagine instead using maximum likelihood to estimate the parameters, the Bayesian framework has several advantages here: for example, it allows for
incorporation of external information (e.g. that most genetic markers will, individually, have small effects), and it takes into account of uncertainty in parameter estimates when making other inferences (e.g.~phenotype prediction).

For the mean $\mu$ and the inverse of error variance, $\tau$, we use the standard
conjugate prior distributions:
\begin{align}
\tau &\sim \mbox{Gamma}(\kappa_1, \kappa_2), \\
\mu|\tau &\sim \mbox{N}(0, \sigma_{\mu}^2\tau^{-1}),
\end{align}
where $\kappa_1$ and $\kappa_2$ denote, respectively, shape and rate parameters of a Gamma distribution. Specifically 
we consider the posterior that arises in the limits $\kappa_1\to 0$, $\kappa_2\to 0$ and $\sigma_{\mu}^2 \to \infty$. These limits correspond to improper priors, but the resulting posteriors are proper, and scale appropriately with shifting or scaling of the phenotype vector $\by$ \cite{Servin:2007}. In particular, these priors have
the property that conclusions will be unaffected by changing the units of measurement of the phenotype, which seems desirable for a method intended for general application. 

Prior specification for the remaining hyper-parameters $\pi, \sigma_a^2, \sigma_b^2$ is perhaps more important. 
Our approach is to extend the prior distributions for BVSR described in \cite{Guan:2011}.

Following \cite{Guan:2011} we place a uniform prior on $\log\pi$:
\begin{equation}
\label{pi_prior}
\log(\pi) \sim \mbox{U}(\log(1/p), \log(1) ),
\end{equation}
where $p$ is total number of markers being analyzed. The upper and lower limits of this prior  were chosen so that $\pi$ (the expected proportion of 
markers with non-zero $\tbeta$) ranges from $1/p$ to $1$.
A uniform prior on $\log\pi$ reflects the fact that uncertainty in $\pi$ in a typical GWAS will span orders of magnitude. A common alternative
(see e.g. \cite{Logsdon:2010, Habier:2011}) is a uniform distribution on $\pi$,
but as noted in \cite{Guan:2011} this puts large weight on large
numbers of markers having non-zero $\tbeta$ (e.g. it would correspond to placing 50\% prior probability to the event that more than half of the markers have non-zero $\tbeta$, and correspond to placing 90\% prior probability to the event that more than 10\% of the markers have
non-zero $\tbeta$). 

To specify priors for $\sigma_a$  and $\sigma_b$, we exploit 
the following idea from \cite{Guan:2011}: prior specification is easier 
if we first re-parameterize the model in terms of more interpretable quantities. Specifically we extend ideas from \cite{Guan:2011} to 
re-parameterize the model in terms of the (expected) proportion of phenotypic variance explained by the sparse effects and by the random effects.

To this end, we define PVE (the total proportion of variance
in phenotype explained by the sparse effects and random effects terms together)
and PGE (the proportion of genetic variance explained by the sparse effects terms)
as functions of $\tbbeta$, $\bu$ and $\tau$:
\begin{align}
\label{PVE_def}
\mbox{PVE}(\tbbeta, \bu, \tau) & :=\frac{\mbox{V}(\bX\tbbeta+\bu)}{\mbox{V}(\bX\tbbeta+\bu)+\tau^{-1}}, \\  
\label{PGE_def}
\mbox{PGE}(\tbbeta, \bu) & :=\frac{\mbox{V}(\bX\tbbeta)}{\mbox{V}(\bX\tbbeta+\bu)}, \\
\intertext{where the function $V(\bx)$ is defined as}  
\label{Vfunc_def}
\mbox{V}({\bf x}) &:=\frac{1}{n}\sum_{i=1}^n (x_i-\overline {x})^2.
\end{align}
These definitions ensure that both PVE and PGE must lie in the interval $[0,1]$. PVE reflects how well one could predict phenotype $\by$ from the available SNPs if one knew the optimal $\tbbeta$ as well as the random effects $\bu$; together with PVE, PGE reflects how well one could predict phenotype $\by$ using $\tbbeta$ alone. 
%Note keen on this phrase - can we avoid adding it? MS
%\textcolor{red}{[Notice that in polygenic modeling, PGE does not reflect all the contribution from the large effect markers, but rather the additional effects of those markers above the polygenic background.]} 

Since PVE and PGE are functions of $(\tbbeta, \bu, \tau)$, whose
distributions depend on hyper-parameters $\pi, \sigma_a,\sigma_b$,
the prior distribution for PVE and PGE depends on the priors assigned to these hyper-parameters. In brief, our aim is to choose priors for the two hyper-parameters $\sigma_a^2$ and $\sigma_b^2$ so that the induced priors on both PVE and PGE are roughly uniform on 0 and 1. (Other distributions could
be chosen if desired, but we consider this uniform distribution one reasonable
default.) However, because the relationship between the distribution of PVE, PGE 
and the hyper-parameters is not simple, we have to make some approximations.

Specifically, we introduce $h,\rho$ as approximations (they are ratios of expectations rather than expectations of ratios) to the expectations of PVE and PGE, respectively:
\begin{align}
\label{h_def}
h   &:=\frac{p\pi s_a\sigma_a^2+s_b\sigma_b^2}{p\pi s_a\sigma_a^2+s_b\sigma_b^2+1}, \\
\label{rho_def}
\rho  &:=\frac{p\pi s_a\sigma_a^2}{p\pi s_a\sigma_a^2+s_b\sigma_b^2},
\end{align}
where $s_a$ is the average variance of genotypes across markers, and $s_b$ is the mean of diagonal elements in $\bK$. In other words, $s_a=\frac{1}{np}\sum_{i=1}^p\sum_{j=1}^n x_{ij}^2$ and $s_b=\frac{1}{n}\sum_{i=1}^n k_{ii}$, where $x_{ij}$ and $k_{ij}$ are the $ij$th elements of matrices $\bX$ and $\bK$, respectively.
See Text S1 Detailed Methods for derivations.
Intuitively, the term $p\pi s_a\sigma_a^2$ captures the expected genetic variance contributed by the sparse effects term (relative to the error variance), because $p\pi$ is the expected number of causal markers, $\sigma_a^2$ is the expected effect size variance of each causal marker (relative to the error variance), and $s_a$ is the average variance of marker genotypes. Similarly, the term $s_b\sigma_b^2$ captures the expected genetic variance contributed by the random effects term (relative to the error variance), because $\sigma_b^2$ is the expected variance of the random effects (relative to the error variance) when the relatedness matrix has unit diagonal elements, while $s_b$ properly scales it when not.

The parameter $\rho$ provides a natural bridge between the LMM and BVSR: when $\rho=0$ BSLMM becomes the LMM,  and when $\rho=1$ BSLMM becomes BVSR. In practice, when the data favors the LMM, the posterior distribution of $\rho$ would mass near 0, and when the data favors BVSR, the posterior distribution of $\rho$ would mass near 1. 

In summary, the above re-parameterizes the model in terms of $(h,\rho,\pi)$ instead
of $(\sigma_a, \sigma_b, \pi)$. Now, instead of specifying prior distributions for $\sigma_a, \sigma_b$, we rather specify prior distributions for $h,\rho$. Specifically we use 
uniform prior distributions for $h,\rho$: 
\begin{align}
\label{h_prior}
h&\sim \mbox{U}(0,1), \\
\label{rho_prior}
\rho&\sim \mbox{U}(0,1),
\end{align}
independent of one another and of $\pi$. Since $h$ and $\rho$ approximate PVE and PGE, these prior distributions should lead to reasonably
uniform prior distributions for PVE and PGE, which we view as reasonable defaults for general use. (If one had specific information about
PVE and PGE in a given application then this could be incorporated here.) In contrast it seems much harder to directly specify reasonable default
priors for $\sigma_a,\sigma_b$ (although these priors on $h,\rho,\pi$ do of course imply priors for $\sigma_a,\sigma_b$; see Text S1 Detailed Methods).

Note that we treat $h$ and $\rho$ as approximations to PVE and PGE only for prior specification; when estimating PVE and PGE from data we do so directly using their definitions (\ref{PVE_def}) and (\ref{PGE_def}) (see below for details).

\subsection*{Posterior Sampling Scheme} 
\label{subsec:comp}
To facilitate computation, we use the widely-used approach from \cite{George:1993} of introducing a vector of binary indicators $\bgamma=(\gamma_1, \cdots, \gamma_p) \in \{0, 1\}^p$ that indicates whether the corresponding coefficients $\tbbeta$ are non-zero.
The point-normal priors for $\tbbeta$ can then be written 
\begin{align}
\gamma_i &\sim \mbox{Bernoulli}(\pi), \\
\tbbeta_{\bgamma} &\sim \mbox{MVN}_{|\bgamma|}(0, \sigma_a^2\tau^{-1}\bI_{|\bgamma|}), \\
\tbbeta_{-\bgamma} &\sim \bdelta_0,
\end{align}
where $\tbbeta_{\bgamma}$ denotes the sub-vector of $\tbbeta$ corresponding to the entries $\{i:\gamma_i=1\}$; $\tbbeta_{-\bgamma}$ denotes the sub-vector of $\tbbeta$ corresponding to the other entries, $\{i: \gamma_i=0\}$; and $|\bgamma|$ denotes the number of non-zero entries in $\bgamma$. 
We use MCMC to obtain posterior samples of ($h, \rho, \pi, \bgamma$) on the product space $(0,1)\times(0,1)\times(0,1)\times\{0,1\}^p$, which is given by
\begin{equation} \label{eqn:posterior}
P(h,\rho, \pi, \bgamma | \by)\propto P(\by|h, \rho, \pi, \bgamma)P(h)P(\rho)P(\gamma|\pi)P(\pi).
\end{equation}
The marginal likelihood $P(\by | h, \rho, \pi, \bgamma)$ can be computed analytically
integrating out $(\tbbeta,\bu,\tau)$; see below for further details. 
We use a Metropolis-Hastings algorithm to draw posterior samples from the above marginal distribution. 
In particular, we use a rank based proposal distribution for $\bgamma$ \cite{Guan:2011}, which focus more of the computational time on examining SNPs with stronger marginal associations. 

We use the resulting sample from the posterior distribution (\ref{eqn:posterior}) to estimate PVE and PGE as follows. For each sampled value of $(h, \rho, \pi, \bgamma)$, we sample a corresponding value for $(\tau,\tbbeta,\bu)$ from the conditional distribution $P(\tau, \tbbeta, \bu | \by, h, \rho, \pi, \bgamma)$. We then use each sampled value of $(\tau,\tbbeta,\bu)$ to compute a sampled value of PVE and PGE, using equations (\ref{PVE_def}) and (\ref{PGE_def}). We estimate the posterior mean and standard deviation of PVE, PGE, from these samples.

The novel part of our algorithm is a new efficient approach to evaluating the likelihood $P(\by | h, \rho, \pi, \bgamma)$ that considerably reduces
the overall computational burden of the algorithm. The direct naive approach to evaluating this likelihood involves a matrix inversion and a matrix determinant calculation that scale cubically with the number of individuals $n$,
and this cost is incurred every iteration as hyper parameter values change. Consequently, this approach is impractical for typical association studies with large $n$, and
{\it ad hoc} approximations are commonly used to reduce the burden.  For example, both \cite{Lee:2008} and \cite{Segura:2012} fix $\sigma_b^2$ to some pre-estimated value. As we show later, this kind of approximation can reduce the accuracy of predicted phenotypes. Here, we avoid such approximations by exploiting recently developed computational tricks for LMMs \cite{Lippert:2011, Zhou:2012}. The key idea is to perform a single eigen-decomposition and use the resulting unitary matrix (consisting of all eigen vectors) to transform both phenotypes and genotypes to make the transformed values follow independent normal distributions. By extending these tricks to BSLMM we evaluate the necessary likelihoods much more efficiently. Specifically, after a single $n^3$ operation at the start of the algorithm, the per iteration computational burden is linear in $n$ (the same as BVSR), allowing large studies to be analyzed.

Full details of the sampling algorithm appear in Text S2. 

\subsection*{URLs}
Software implementing our methods is included in the GEMMA software package, which is freely available at \url{http://stephenslab.uchicago.edu/software.html}.

% Results and Discussion can be combined.
\clearpage
\newpage
\section*{Results}
\subsection*{Simulations: PVE Estimation}
Both the LMM and BVSR have been used to estimate the PVE \cite{Yang:2010, Guan:2011}. 
Since the LMM assumes that all SNPs have an effect, while BVSR assumes that only a small proportion of SNPs have an effect, we hypothesize that BVSR will perform better when the true underlying genetic structure is sparse and LMM will perform better when the true genetic structure is highly polygenic. Further, because BSLMM includes both as special cases, we hypothesize that BSLMM will perform well in either scenario. 
 
To test these hypotheses, we perform a simulation using real genotypes at about 300,000 SNPs in 3,925 Australian individuals \cite{Yang:2010}, and simulate phenotypes under two different scenarios. In Scenario I we simulate a fixed number $S$ of causal SNPs (with $S=10,100,1000,10000$), with effect sizes coming from a standard normal distribution. These simulations span a range of genetic architectures, from very sparse to highly polygenic. In Scenario II we simulate two groups of causal SNPs, the first group containing a small number of SNPs of moderate effect ($S=10$ or $S=100$),
plus a second larger group of $10,000$ SNPs of small effect representing a ``polygenic component".
This scenario might be considered more realistic, containing a mix of small and larger effects.
For both scenarios we added normally-distributed errors to phenotype to create datasets with true PVE =0.6 and 0.2 (equation \ref{PVE_def}). We simulate 20 replicates in each case, and run the algorithms with all SNPs, including the simulated causal variants,
so that the true PVE for typed markers is either 0.6 or 0.2 (if we excluded the causal variants then the true PVE would be unknown).

\begin{figure}[!ht]
\centering
\subfigure[RMSE, true PVE=0.2]{
\includegraphics[width=17.35cm,keepaspectratio=true]{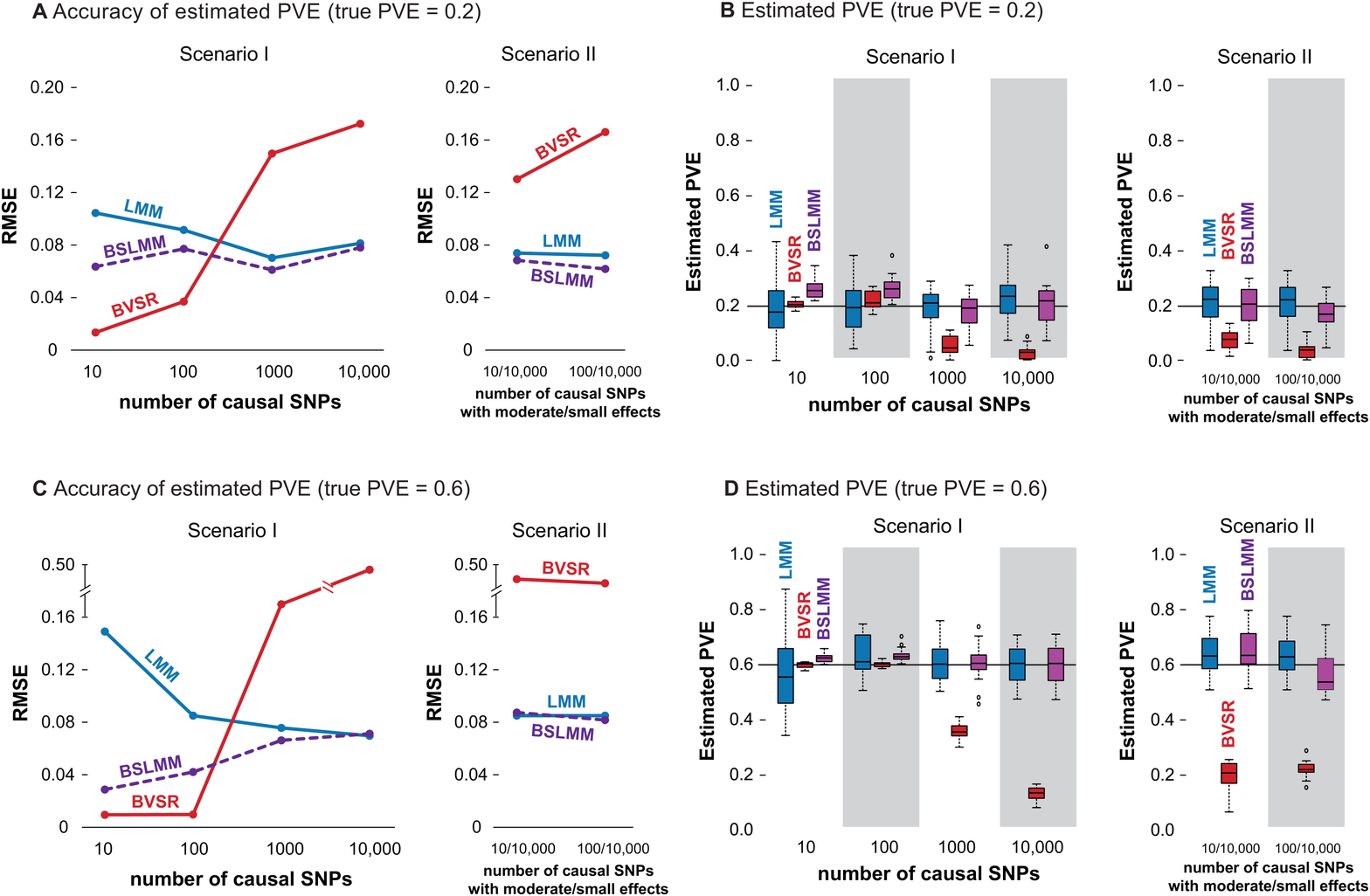}
%\includegraphics[width=200pt, height=200pt, angle=0]{Figures/PVE_oz3925_bimbam_pve2_rmse.png}
%\includegraphics[width=7.6cm,keepaspectratio=true]{Figures/True_PVE_02_RMSE.eps}
%  \label{fig1:a}
% }
% \subfigure[PVE Estimates, true PVE=0.2]{
  %\includegraphics[width=200pt, height=200pt, angle=0]{Figures/PVE_oz3925_bimbam_pve2_h.png}
%  \includegraphics[width=7.6cm,keepaspectratio=true]{Figures/True_PVE_02_PVE.eps}
%  \label{fig1:b}
% }
% \subfigure[RMSE, true PVE=0.6]{
 %\includegraphics[width=200pt, height=200pt, angle=0]{Figures/PVE_oz3925_bimbam_pve6_rmse.png}
% \includegraphics[width=7.6cm,keepaspectratio=true]{Figures/True_PVE_06_RMSE.eps}
%  \label{fig1:c}
% }
% \subfigure[PVE Estimates, true PVE=0.6]{
%\includegraphics[width=200pt, height=200pt, angle=0]{Figures/PVE_oz3925_bimbam_pve6_h.png}
%  \includegraphics[width=7.6cm,keepaspectratio=true]{Figures/True_PVE_06_PVE.eps}
%  \label{fig1:d}
 }
\caption{Comparison of PVE estimates from LMM (blue), BVSR (red) and BSLMM (purple) in two simulation scenarios. The x-axis show the number of causal SNPs (Scenario I) or the number of medium/small effect SNPs (Scenario II). Results are based on 20 replicates in each case. (A) (true PVE=0.2) and (C) (true PVE=0.6) show RMSE of PVE estimates. (B) (true PVE=0.2) and (D) (true PVE=0.6) show boxplots of PVE estimates, where the true PVE is shown as a horizontal line. Notice a break point on the y-axis in (C).}
\label{fig:sim_pve}
\end{figure}

Figures \ref{fig:sim_pve}A and \ref{fig:sim_pve}C, show the root of mean square error (RMSE) of the PVE estimates obtained by each method, and Figure \ref{fig:sim_pve}B and \ref{fig:sim_pve}D summarize the corresponding distributions of PVE estimates. In agreement with our original hypotheses, BVSR performs best (lowest RMSE) when the true model is sparse (e.g.~Scenario I, $S= 10$ or $S= 100$ in Figures \ref{fig:sim_pve}A, \ref{fig:sim_pve}C). However, it performs very poorly under all the other, more polygenic, models. 
This is due to a strong downward bias in its PVE estimates (Figures \ref{fig:sim_pve}B, \ref{fig:sim_pve}D).
Conversely, under the same scenarios, LMM is the least accurate method. This is because the LMM estimates have much larger variance than
the other methods under these scenarios (Figure \ref{fig:sim_pve}B,\ref{fig:sim_pve}D), although, interestingly, LMM is approximately unbiased
even in these settings where its modeling assumptions are badly wrong.
%\textcolor{red}{[Despite its poor performance in the sparse scenarios, it is worth noting that the estimates from LMM are approximately
%unbiased.]}
 As hypothesized, BSLMM is robust across a wider range of settings than the other methods: its performance lies between LMM and BVSR when the true model is sparse, and provides similar accuracy to LMM when not. 
   
Of course, in practice, one does not know in advance the correct
genetic architecture. This makes the stable performance of BSLMM 
across a range of settings very appealing. Due to the poor
performance of BVSR under highly polygenic models, we would not now recommend it
for estimating PVE in general, despite its good performance when its assumptions are met.

\subsection*{Simulations: Phenotype Prediction}

We also compare the three methods on their ability to predict phenotypes from genotypes,
using the same simulations. 

To measure prediction performance, we use relative prediction gain (RPG; see Text S1 Detailed Methods). In brief, RPG is a standardized version of mean square error: RPG=1 when accuracy is as good as possible given the simulation setup, and RPG=0 when accuracy is the same as simply predicting everyone to have the mean phenotype value. RPG can be negative if accuracy is even worse than that. 

\begin{figure}[!ht]
\centering
\includegraphics[width=8.3cm,keepaspectratio=true]{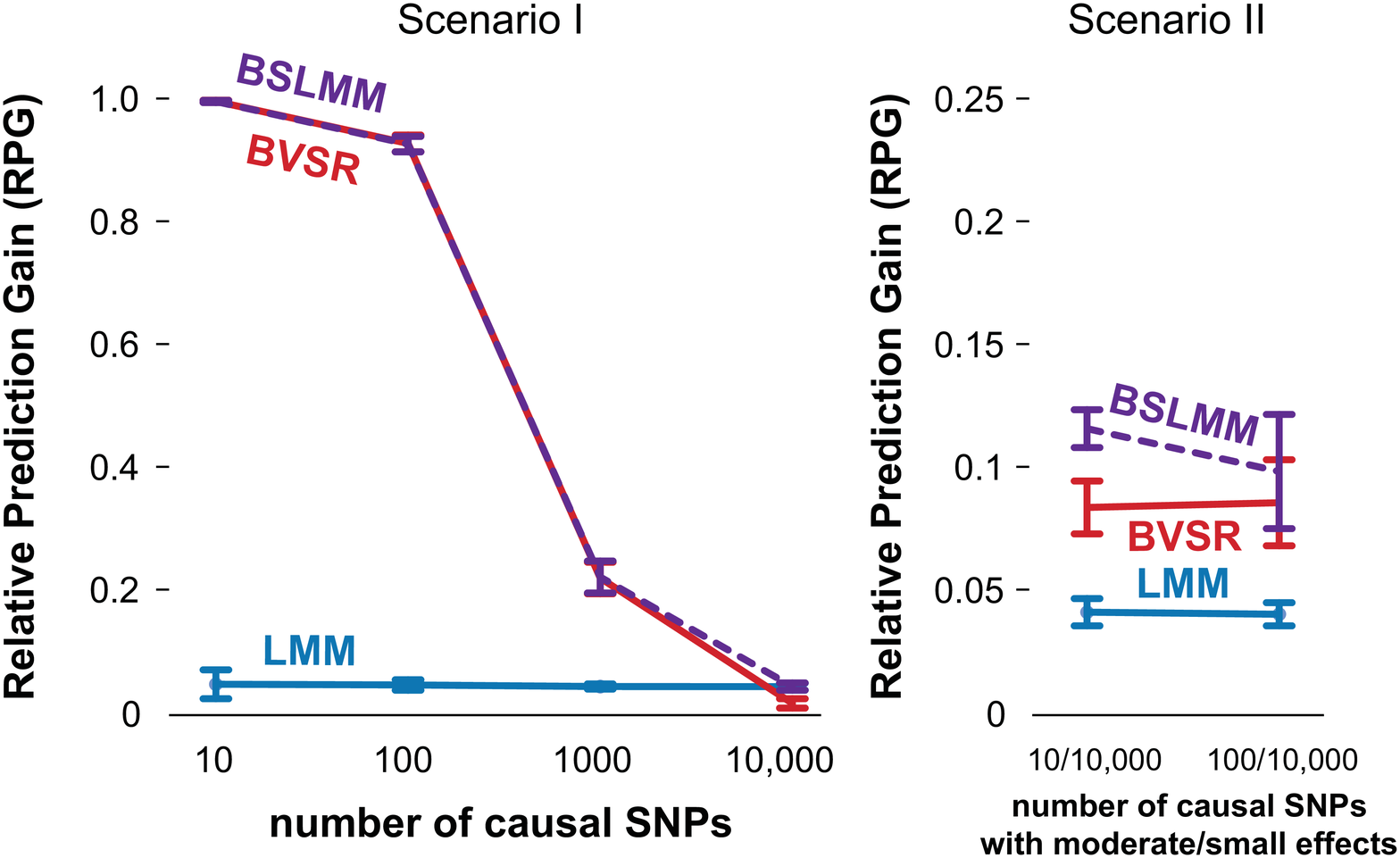}
\caption{Comparison of prediction performance of LMM (blue), BVSR (red) and BSLMM (purple) in two simulation scenarios, where all causal SNPs are included in the data. Performance is measured by Relative Predictive Gain (RPG). True PVE=0.6. Means and standard deviations (error bars) are based on 20 replicates. The x-axis show the number of causal SNPs (Scenario I) or the number of medium/small effect SNPs (Scenario II). }
\label{fig:sim_pred_rpg}
\end{figure}

Figure \ref{fig:sim_pred_rpg} compares RPG of different methods for simulations with PVE=0.6 (results for PVE=0.2 are qualitatively
similar, not shown). Interestingly, for phenotype prediction, the relative performance of the methods differs from results for PVE estimation. In particular, LMM performs poorly compared with the other two methods in all situations, except for Scenario I with $S=10,000$, the Scenario that comes closest to matching the underlying assumptions of LMM. As we expect,  BSLMM performs similarly to BVSR in scenarios involving smaller numbers of causal SNPs (up to $S=1,000$ in Scenario I), and outperforms it in more polygenic scenarios involving large numbers of SNPs of small effect
(e.g.~Scenario II). This is presumably due to the random effect in BSLMM that captures the polygenic component, or, equivalently, due to the mixture of two normal distributions in BSLMM that better captures the actual distribution of effect sizes. The same qualitative patterns hold when we redo these simulation comparisons excluding the causal SNPs (Figure \ref{fig:sim_pred_exclude_rpg}) or
use correlation instead of RPG to assess performance 
(Figure \ref{fig:sim_pred_cor} and \ref{fig:sim_pred_exclude_cor}).

For a potential explanation why LMM performs much less well for phenotype 
prediction than for PVE estimation, 
we note the difference between these two problems: to accurately estimate PVE 
it suffices to estimate the ``average" effect size reliably, whereas accurate phenotype prediction requires accurate estimates of individual effect sizes. In situations
where the normal assumption on effect sizes is poor, LMM tends to considerably underestimate
the number of large effects, and overestimate the number of small effects.
These factors can cancel one another out in PVE estimation, but both tend to reduce
accuracy of phenotype prediction.

\subsection*{Estimating PVE in Complex Human Traits}
To obtain further insights into differences between LMM, BVSR and BSLMMM, we apply
all three methods to estimate the PVE for five traits in two human GWAS data sets. The first data set contains height measurements of 3,925 Australian individuals with about 300,000 typed SNPs. The second data set contains four blood lipid measurements, including high-density lipoprotein (HDL), low-density lipoprotein (LDL), total cholesterol (TC) and triglycerides (TG) from 1,868 Caucasian individuals with about 550,000 SNPs. The narrow sense heritability for height is estimated to be 0.8 from twin-studies \cite{Macgregor:2006, Yang:2010}. The narrow sense heritabilities for the lipid traits have been estimated, in isolated founder populations, to be 0.63 for HDL, 0.50 for LDL, 0.37 for TG in Hutterites \cite{Abney:2001}, and 0.49 for HDL, 0.42 for LDL, 0.42 for TC and 0.32 for TG in Sardinians \cite{Pilia:2006}. 

Table \ref{tab:real_pve} shows PVE estimates from the three methods for the five traits. PVE estimates from BVSR are consistently much smaller than those obtained by LMM and BSLMM, which are almost identical for two traits and similar for the others. Estimates of PVE from both LMM and BSLMM explain over 50\% of the narrow sense heritability of the five traits, suggesting that a sizable proportion of heritability of these traits can be explained, either directly or indirectly, by available SNPs. 

These results, with LMM and BSLMM providing similar estimates of PVE, and estimates from BVSR being substantially
lower, are consistent with simulation results for a trait with substantial polygenic component.
One feature of BSLMM, not possessed by the other two methods, is that it can be used 
to attempt to quantify the relative contribution of a polygenic component, through estimation of
PGE,  which is the proportion of total genetic variance explained by ``large" effect size SNPs (or more precisely, by the additional effects of those SNPs above a polygenic background).
Since the PGE is defined within an inevitably over-simplistic model, specifically that effect sizes
come from a mixture of two normal distributions, and also because it will be influenced
by unmeasured environmental factors that correlate with genetic factors, we caution against over-interpreting the estimated values. We also note that estimates of PGE for these data (Table \ref{tab:real_pve}) are generally not very precise (high posterior standard deviation). 
Nonetheless, it is interesting that the estimated PGE for height, at 0.12,
is lower than for any of the lipid traits (ranging from 0.18 for TG to 0.46 for TC), and 
that all these estimates suggest a substantial contribution from small polygenic effects in all five traits.

\subsection*{Predicting Disease Risk in the WTCCC Data Set}
To assess predictive performance on real data, we turn to the Wellcome trust case control consortium (WTCCC) 1 study \cite{WTCCC:2007}, which have been previously used for assessing risk prediction \cite{Evans:2009, Wei:2009, Kooperberg:2010}. These data include about 14,000 cases from seven common diseases and about 3,000 shared controls, typed at a total of about 450,000 SNPs. The seven common diseases are bipolar disorder (BD), coronary artery disease (CAD), Crohn's disease (CD), hypertension (HT), rheumatoid arthritis (RA), type 1 diabetes (T1D) and type 2 diabetes (T2D). 

We compared the prediction performance of LMM, BVSR and BSLMM for all seven diseases. Following \cite{Wei:2009}, we randomly split the data for each disease into a training set (80\% of individuals) and a test set (remaining 20\%), performing 20 such splits for each disease. We estimated parameters from the training set by treating the binary case control labels as quantitative traits, as in \cite{Kang:2010, Zhou:2012}. [This approach can be justified by recognizing the linear model as a first order Taylor approximation to a generalized linear model; we discuss directly modeling binary phenotypes in the Discussion section.] We assess prediction performance in the test set by area under the curve (AUC) \cite{Wray:2010}.

\begin{figure}[!ht]
\begin{center}
\includegraphics[width=17.35cm,keepaspectratio=true]{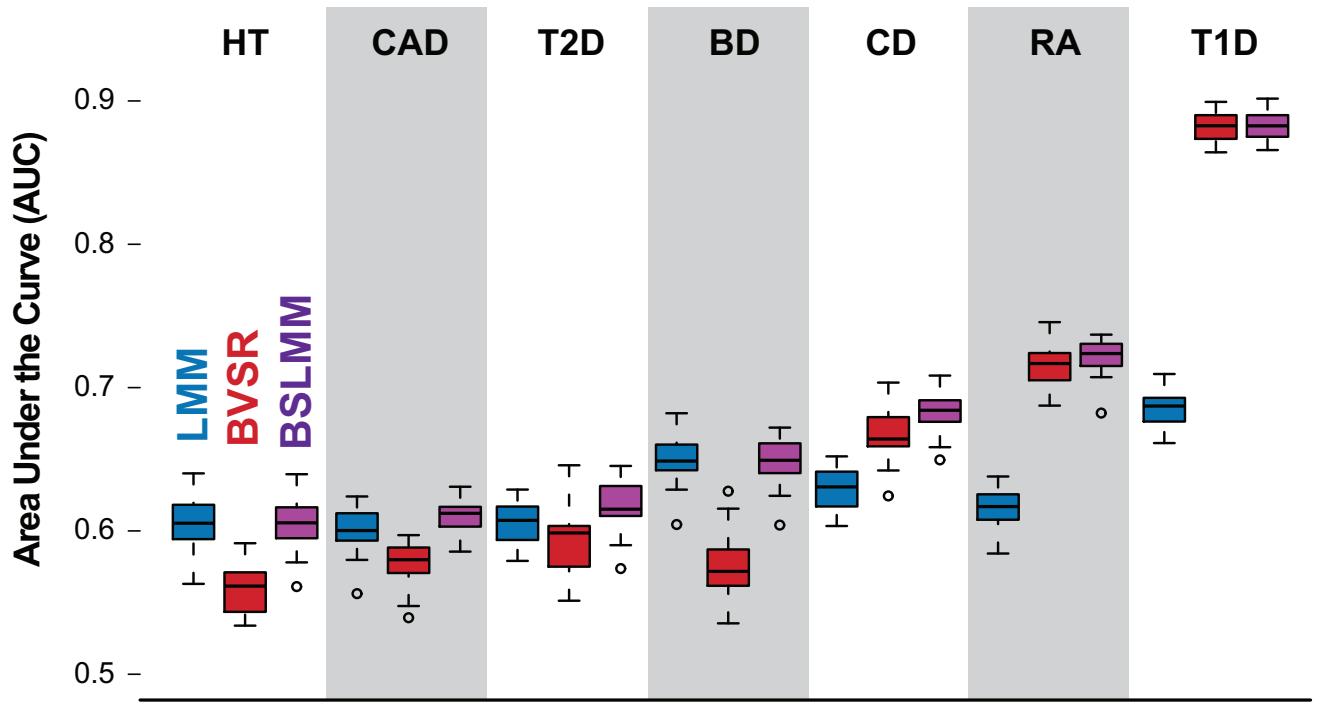}
\end{center}
\caption{Comparison of prediction performance of LMM (blue), BVSR (red) and BSLMM (purple) for seven diseases in the WTCCC data set. Performance is measured by area under the curve (AUC), where a higher value indicates better performance. The order of the diseases is based on the performance of BSLMM. The mean and standard deviation of AUC scores for BSLMM in the seven diseases are 0.60 (0.02) for HT, 0.60 (0.03) for CAD, 0.61 (0.03) for T2D, 0.65 (0.02) for BD, 0.68 (0.02) for CD, 0.72 (0.01) for RA, 0.88 (0.01) for T1D.}
\label{fig:real_pred_wtccc_auc}
\end{figure}

Figure \ref{fig:real_pred_wtccc_auc} shows AUC for the three methods on all seven diseases. As
in our simulations, we find BSLMM performs as well as or better than either of the other two methods for all seven diseases. Indeed, the performance of BSLMM appears to compare favorably with previous methods applied to the same data set \cite{Evans:2009, Wei:2009, Kooperberg:2010} (a 
precise comparison with previous results is difficult, as some studies use slightly different splitting strategies \cite{Evans:2009, Kooperberg:2010} and some do not perform full cross validation \cite{Wei:2009}). As might be expected from the simulation results, 
BVSR performs better than LMM in diseases
where a small number of relatively strong associations were identified
in the original study  \cite{WTCCC:2007} (CD, RA and T1D) and worse in the others. We obtained qualitatively similar results
when we measured performance using the Brier score instead of AUC (Text S3; Figure \ref{fig:real_pred_wtccc_brier}).

Finally, we caution that, although BSLMM performs well here relative to other methods, at the present time, for these diseases, its prediction accuracy is unlikely to be
of practical use in human clinical settings. In particular, in these simulations the number of cases and
controls in the test set is roughly equal, which represents a much easier problem than clinical settings where disease prevalence is generally low even for common diseases (see \cite{Wei:2009} for a relevant discussion).

\subsection*{Predicting Quantitative Phenotypes in Heterogeneous Stock Mice}
In addition to the WTCCC data set, we also assess perdition performance using a mouse data set \cite{Valdar:2006}, which has been widely used to compare various phenotype prediction methods \cite{Lee:2008, Legarra:2008, Campos:2009}. The mouse data set is substantially smaller than the human data ($n\approx 1000,p\approx 10,000$, with exact numbers varying slightly depending on the phenotype and the split). This makes it computationally feasible to compare with a wider range of other methods. Therefore, we include in our comparisons here five other related approaches, some of which have been proposed previously for phenotype prediction. 
Specifically we compare with: 
\begin{enumerate}
\item {\bf LMM-Bayes}, a Bayesian version of the LMM, which we fit by setting $\rho=0$ in BSLMM using our software. 
\item {\bf Bayesian Lasso} \cite{Park:2008, Campos:2009}, implemented in the R package BLR \cite{Campos:2009}. 
\item {\bf BayesA-Flex}, our own modification of BayesA, which assumes a $t$ distribution for the effect sizes. Our modification involves
estimating the scale parameter associated with the $t_4$ distribution from the data (Text S1 Detailed Methods). Although the original BayesA performs poorly in this data set \cite{Legarra:2008}, this simple modification greatly improves its prediction performance (emphasizing again the general importance of estimating hyper-parameters from the data rather than fixing them to arbitrary values). We modified the R package BLR \cite{Campos:2009} to obtain posterior samples from this model. 
\item{\bf  BayesC$\pi$,} implemented in the online software GenSel  \cite{Habier:2011}. This implementation does not allow random effects, and therefore uses the same model as our BVSR, although with different prior distributions.
\item {\bf BSLMM-EB} (where EB stands for empirical Bayes\footnote{This is a slight abuse of terminology, since in this method the estimated hyper parameters are obtained under the model with $\tbbeta=0$}), an approximation method to fit BSLMM. The method fixes the variance component $\sigma_b^2$ to its REML (restricted maximum likelihood) estimate obtained with $\tbbeta=0$, which is one of several strategies used  in previous studies to alleviate the computational burden of fitting models similar to BSLMM \cite {Lee:2008, Segura:2012}. We sample posteriors from this model using our software. 
\end{enumerate}
See Text S1 Detailed Methods for further details. 

Following previous studies that have used these data for prediction
\cite{Lee:2008, Legarra:2008, Campos:2009}
we focused on three quantitative phenotypes: CD8, MCH and BMI. These phenotypes  have very different estimated narrow sense heritabilities:  0.89, 0.48, and 0.13 respectively \cite{Valdar:2006b}. Table \ref{tab:real_pve_mouse} lists estimates of some key quantities for the three traits -- including PVE, PGE and $\log_{10}(\pi)$ -- obtained from LMM, BVSR and BSLMM. All three methods agree well on the PVE estimates, suggesting that the data is informative enough to overwhelm differences in prior specification for PVE estimation.

Following  \cite{Lee:2008, Legarra:2008}, we divide the mouse data set roughly half and half into a training set and a test set. As the mice come from 85 families, and individuals within a family are more closely related than individuals from different families, we also follow previous studies and use two different splits of the data: the {\it intra-family} split mixes all individuals together and randomly divides them into two sets of roughly equal size; the {\it inter-family} split randomly divides the 85 families into two sets, where each set contains roughly half of the individuals. We perform 20 replicates for each split of each phenotype. It is important to note that the intra-family split represents an easier setting for phenotype prediction, not only because individuals in the test set are more related genetically to those in the training set, but also because the individuals in the test set often share a similar environment with those in the training set (specifically, in the intra-family split, many individuals in the test set share a cage with individuals in the training set, but this is not the case in the inter-family split).

%In particular, animals in this data set are housed in different cages and those reside in a same cage share a common environment. The shared environment could contribute to prediction accuracy, if both the training set and the test set contain animals from the same cage. This is indeed the case for the intra-family split, but not the case for the inter-family split. Specifically, in the intra-family split, 312 (min 296, max 330) out of 423 cages are shared between the training set and the test set for CD8, 344 (min 335, max 363) out of 450 cages are shared for MCH, and 394 (min 375, max 410) out of 526 cages are shared for BMI. While in the inter-family split, only 4 (min 1, max 7) cages are shared between the training set and the test set for CD8, 4 (min 1, max 7) cages are shared for MCH and 7 (min 3, max 12) cages are shared for BMI. Therefore, the inter-family split effectively disentangles the confounding between environmental effects (i.e. cage) and genetic effects (i.e. family), and abolishes the potential influence from the cage effects to our assessment of prediction accuracy of various methods.

We apply each method using genotypes only, without other covariates. We obtain effect size estimates in the training data set, and assess prediction performance using these estimates in the test set by root of mean square error (RMSE), where the mean is across individuals in the test set.  We contrast the performance of other methods to BSLMM by calculating the RMSE difference, where a positive number indicates worse performance than BSLMM. We perform 20 inter-family splits and 20 intra-family splits for each phenotype.

\begin{figure}[!ht]
\begin{center}
\includegraphics[width=16cm,keepaspectratio=true]{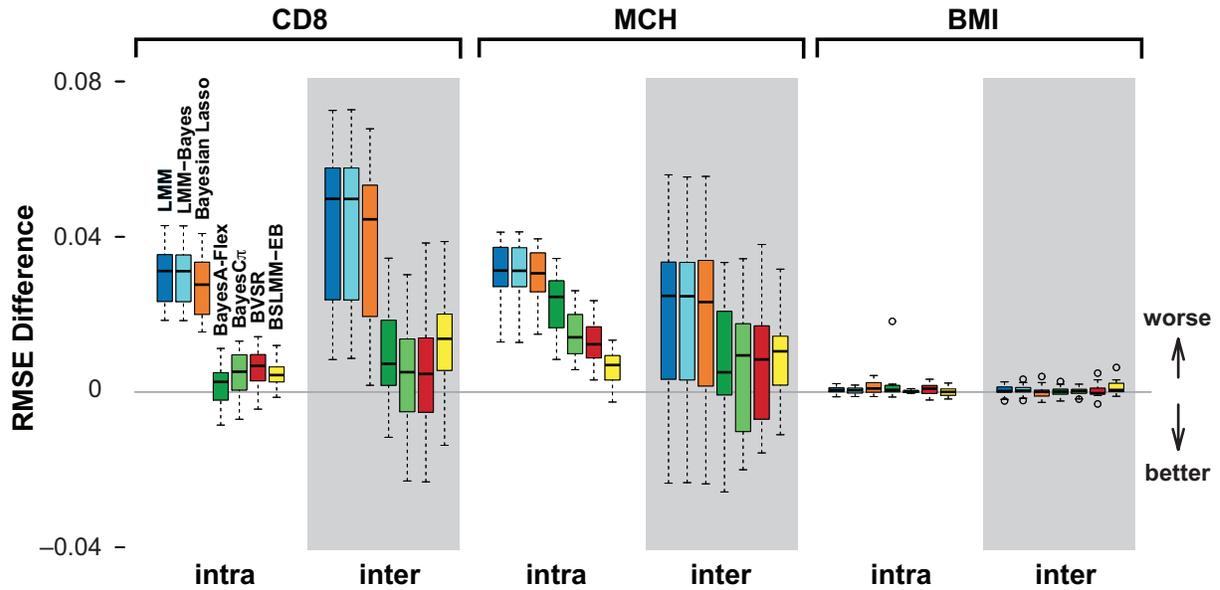}
\end{center}
\caption{Comparison of prediction performance of several models with BSLMM for three traits in the heterogenous stock mouse data set. Performance is measured by RMSE difference with respect to BSLMM, where a positive value indicates worse performance than BSLMM. The x-axis shows two different ways to split the data into a training set and a test set, each with 20 replicates. The mean RMSE of BSLMM for the six cases are 0.70, 0.80, 0.79, 0.90, 0.98 and 0.99, respectively. }
\label{fig:real_pred_mouse_rmse}
\end{figure}

Figure \ref{fig:real_pred_mouse_rmse} summarizes the prediction accuracy, measured by RMSE, of each method compared against BSLMM. Measuring prediction performance by correlation gives similar results (Figure \ref{fig:real_pred_mouse_cor}). For the low-heritability trait BMI, where no measured SNP has a large effect, all methods perform equally poorly. For both the more heritable traits, CD8 and MCH, BSLMM consistently outperformed all other methods, which seem to split into two groups: LMM, LMM-Bayes and Bayesian Lasso perform least well and similarly to one another on average; BVSR, BayesA-Flex, BayesC$\pi$ and BSLMM-EB perform better, and similarly to one another on average. A general trend here is that accuracy tends to increase as model assumptions improve in their ability to capture both larger genetic effects, and the combined ``polygenic" contribution of smaller genetic effects (and possibly also confounding environmental effects correlating with genetic background). In particular, the $t_4$ distribution
underlying BayesA-Flex, which has a long tail that could capture large effects, performs noticeably better than either the normal
or double-exponential distributions for effect sizes underlying LMM and Bayesian Lasso.  

Comparisons of pairs of closely-related methods yield additional insights into factors that do and do not affect
prediction accuracy. The fact that BSLMM performs better than BSLMM-EB illustrates how the approximation used in BSLMM-EB can degrade prediction accuracy, and thus demonstrates the practical benefits
of our novel computational approach that avoids this approximation. Similarly, the superior performance of BayesA-Flex over BayesA (which performed poorly; not shown) also illustrates the benefits of estimating hyper parameters from the data, rather than fixing them to pre-specified values.
The similar performance between BVSR and BayesC$\pi$, which fit the same model but with different priors, suggests that,
for these data, results are relatively robust to prior specification. Presumably, this is because the data
are sufficiently informative to overwhelm the differences in prior.

\subsection*{Computational Speed}

The average computational time taken for each method on the Mouse data is shown in Table \ref{tab:comptime_mouse}. Some 
differences in computational time among methods may reflect implementational issues, including the language environment in which the methods are implemented, rather than fundamental differences between algorithms. 
In addition, computing times for many methods will be affected by the number of iterations used, and we did not undertake a comprehensive evaluation of how many iterations suffice for each algorithm. Nonetheless, the results indicate that
our implementation of BSLMM is competitive in computing speed with the other (sampling-based) implementations considered here.

In particular, we note that BSLMM is computationally faster than BVSR. This is unexpected, since BSLMM is effectively BVSR plus a random effects term, and the addition of a random effects term usually complicates computation.  The explanation
for this is that the (per-iteration) computational complexity of both BSLMM and BVSR depends, quadratically, on the number of selected SNPs in the sparse effects term ($|\bgamma|$), and this number can be substantially larger with BVSR than with BSLMM, because with BVSR additional SNPs are included to mimic the effect of the random effects term in BSLMM. The size of this effect will vary among data sets, but it can be substantial, particularly  in cases where there are a large number of causal SNPs  with small effects.  

To illustrate this, Table \ref{tab:comptime} compares mean computation time for BSLMM vs BVSR for all data sets used here.  For simulated data with a small number of causal SNPs, BSLMM and BVSR have similar computational times. However, in other cases  (e.g. PVE=0.6, S=10,000 in Scenario I) BSLMM can be over an order of magnitude faster than BVSR. In a sense, this speed improvement of BSLMM over BVSR is consistent with its hybrid nature: in highly polygenic traits, BSLMM tends to behave similarly to LMM, resulting in a considerable speed gain.

\clearpage
\newpage
\section*{Discussion}
% You may title this section "Methods" or "Models". 
% "Models" is not a valid title for PLoS ONE authors. However, PLoS ONE
% authors may use "Analysis" 
We have presented novel statistical and computational methods for 
BSLMM, a hybrid approach for polygenic modeling for GWAS data that simultaneously allows for both a small number of individually large genetic effects, and combined effects of many small genetic effects, with the balance between the two being inferred from the data in hand. This hybrid approach
is both computationally tractable for moderately large data sets 
(our implementation can handle at least 10,000 individuals with 500,000 SNPs on our moderately-equipped modern desktop computer), and is
sufficiently flexible to perform well in a wide range of settings. 
In particular, depending on the genetic architecture, BSLMM is 
either as accurate, or more accurate, than the widely-used LMM
for estimating PVE of quantitative traits. And for phenotype prediction
BSLMM consistently outperformed a range of other approaches on the examples we considered here. By generalizing two widely-used models, and including both as special cases, BSLMM should have many applications beyond polygenic modeling. Indeed, despite its increased computational burden, we believe that BSLMM represents an attractive alternative to the widely-used LASSO \cite{Tibshirani:1996} for general regression-based prediction problems.

Although it was not our focus here, BSLMM can be easily modified to 
analyze binary phenotypes, including for example, a typical human case-control GWAS.  For PVE estimation, one can directly apply BSLMM, treating the 1/0 case-control status as a quantitative outcome, and then apply a
correction factor derived by \cite{Lee:2011} to transform this estimated PVE
on the ``observed scale" to an estimated PVE on a latent liability scale. This correction, for which we supply an alternative derivation in Text S3, 
corrects for both ascertainment and the binary nature of case-control data.
For phenotype prediction, one can again directly apply BSLMM, treating the 1/0 case-control status as a quantitative outcome, and interpret the 
resulting phenotype predictions as the estimated probability of being a case.
Although in principle one might hope to improve on this by modifying
BSLMM to directly model the binary outcomes, using a probit link for example, we have implemented this probit approach and found that not only is it substantially more computationally expensive (quadratic in $n$ instead of linear in $n$), but it performed
slightly worse than treating the binary outcomes as quantitative, at least
in experiments based on the mouse phenotypes considered here (Text S3). This may partly reflect inaccuracies introduced due to the greater computational burden.

The computational innovations we introduce here, building on work by \cite{Lippert:2011, Zhou:2012}, make BSLMM
considerably more tractable than it would otherwise be. Nonetheless, the
computational burden, as with other posterior sampling based methods, remains heavy, both due to memory requirements
(e.g.~to store all genotypes) and CPU time (e.g.~for the large number of sampling iterations required for reasonable convergence). Although more substantial computational resources will somewhat increase the size of data that can be tackled, further methodological innovation will likely be required to apply BSLMM to the very large data sets that are currently being collected. 

In addition to providing a specific implementation that allows BSLMM
to be fitted to moderately large data sets, we hope that our work also helps highlight some more general principles for improving polygenic modeling
methodology. These include:
\begin{enumerate}
\item The benefits of characterizing different methods by their effect size distribution assumptions. While this point may seem obvious, and is certainly not new (e.g. \cite{Goddard:2009b, Hayes:2010}), polygenic modeling applications often focus on the algorithm used to fit the model, rather than the effect size distribution used. While computational issues are very important, and often interact with modeling assumptions, we believe it is important to distinguish, conceptually, between the two. One benefit of characterizing methods by their modeling assumptions is that it
becomes easier to predict which methods will tend to do well in which settings. 
\item The importance of selecting a sufficiently flexible distribution for effect sizes. The version of BSLMM we focus on here  (with $\bK=\frac{1}{p}\bX\bX^T$) assumes a mixture of two (zero-mean) normals for the effect size distribution. Our examples demonstrate the gain in performance this achieves compared to less flexible distributions such as a single normal (LMM) or a point-normal (BVSR). More generally, in our phenotype
prediction experiments, methods with more flexible effect size distributions tended to perform better than those with less flexible distributions.
\item The importance of estimating hyper-parameters from data, rather than fixing them to pre-specified values. Here we are echo-ing and reinforcing 
similar themes emphasized by \cite{Habier:2011} and \cite{Guan:2011}.
Indeed, our comparison between BSLMM and BSLMM-EB for
phenotype prediction illustrates the 
benefits not only of estimating hyper-parameters from the data, but of doing so in an integrated way, rather than as a two-step procedure.
\item The attractiveness of specifying prior distributions for hyper-parameters by focusing on the proportion of variance in phenotypes explained by different genetic components (e.g.~PVE and PGE in our notation). This idea is not limited to BSLMM, and could be helpful even with methods that make use of other effect size distributions.
\end{enumerate}

One question to which we do not know the answer is how often the
mixture of two normal distributions underlying BSLMM will be sufficiently flexible to  capture the actual effect size distribution, and to what extent more flexible distributional assumptions
(e.g. a mixture of more than two normals, or a mixture of $t$ distributions with degrees of freedom estimated from the data) will produce meaningful gains in performance. It seems likely that, at least in some cases, use of a more
flexible distribution will improve performance, and would therefore be preferable if it could be accomplished with reasonable computational expense.
Unfortunately some of the tricks we use to accomplish computational gains here 
may be less effective, or difficult to apply, for more flexible distributions. In particular, the tricks we use from \cite{Lippert:2011} and \cite{Zhou:2012} 
may be difficult to extend to allow for mixtures with more than two components.
In addition, for some choices of effect size distribution, one might
have to perform MCMC sampling on the effect sizes $\tbbeta$ directly, 
rather than sampling $\bgamma$, integrating $\tbbeta$ out analytically, as we do here. It is unclear whether this will necessarily result in a loss of computational efficiency: sampling $\tbbeta$ reduces computational expense per update at the cost of increasing the number of updates necessary (sampling $\bgamma$ by integrating over $\tbbeta$ analytically ensures faster mixing and convergence \cite{George:1997, OHara:2009}).
Because of these issues, it is difficult to predict which effect size distributions will
ultimately provide the best balance between modeling accuracy and computational burden.
Nonetheless, compared with currently available alternatives, we believe that
BSLMM strikes an appealing balance between flexibility, performance, and
computational tractability.

% Do NOT remove this, even if you are not including acknowledgments
\clearpage
\newpage
\section*{Acknowledgments}
We thank Peter Visscher for making the Australian height data available to us. We thank Yongtao Guan for the source code of the software piMASS. We thank Ida Moltke and John Hickey for helpful comments on the manuscript. This research was supported by NIH grant HG02585 to MS, NIH grant HL092206 (PI Y Gilad) and a cross-disciplinary postdoctoral fellowship from the Human Frontiers Science Program to PC. This study also makes use of data generated by the WTCCC. A full list of the investigators who contributed to the generation of the data is available from the WTCCC website. Funding for the WTCCC project was provided by the Wellcome Trust (award 085475).

%\section*{References}
% The bibtex filename
\bibliography{bslmm_plos}

\clearpage
\newpage

\section*{Text S1 Detailed Methods}
\subsection*{GWAS Datasets}
We used four GWAS data sets in the present study.

The first data set contains height measurements for 3925 Australian individuals \cite{Yang:2010}. We used the data set for simulation and PVE estimation. The phenotypes were already regressed out of age and sex effects, and were quantile normalized to a standard normal distribution afterwards \cite{Yang:2010}. A total of 294,831 SNPs were available after stringent quality control \cite{Yang:2010}. We imputed missing SNPs using IMPUTE2 \cite{Howie:2009}, and further excluded five SNPs that have minor allele frequencies below 1\% after imputation. 

The second data set consists of blood lipid measurements for 1868 individuals \cite{Barber:2010}. We used this data for PVE estimation. The individuals came from two study groups: the Cholesterol and Pharmacogenetics (CAP) group \cite{Simon:2006} and the Pravastatin Inflammation/CRP Evaluation (PRINCE) group \cite{Albert:2001}. The PRINCE study consists of two cohorts: one contains individuals with history of cardiovascular diseases (CVD) and the other contains individuals with no history of CVD. From both study groups, we selected all 1868 individuals that have complete low-density lipoprotein (LDL) subfraction measurements. We selected four different blood lipid measurements as phenotypes in the present study: LDL, high-density lipoprotein (HDL), total cholesterol (TC) and triglycerides (TG). Phenotypes were quantile-normalized to a standard normal distribution within each group, corrected for covariates including BMI (body mass index), age, sex, and smoking status effects, and quantile-normalized again \cite{Barber:2010}. Individuals were typed on two different SNP arrays (Illumina HumanHap300 and HumanQuad610 bead chips, Illumina, San Diego, CA). We used all SNPs that appeared in either of the arrays and we imputed missing genotypes using IMPUTE2 \cite{Howie:2009}. We obtained a total of 582,962 SNPs and we used 555,601 polymorphic SNPs with minor allele frequency above 1\% for analysis. 

The third data set is from the Wellcome trust case control consortium (WTCCC) 1 study \cite{WTCCC:2007}. We used this data set to assess phenotype prediction performance. The data set consists of about 14,000 cases of seven common diseases, including 1868 cases of bipolar disorder (BD), 1926 cases of coronary artery disease (CAD), 1748 cases of Crohn's disease (CD), 1952 cases of hypertension (HT), 1860 cases rheumatoid arthritis (RA), 1963 cases of type 1 diabetes (T1D) and 1924 cases of type 2 diabetes (T2D), as well as 2938 shared controls. We obtained quality controlled genotypes from WTCCC and we further imputed missing genotypes using BIMBAM \cite{Guan:2008}, which resulted in a total of 458,868 shared SNPs. All polymorphic SNPs with minor allele frequency above 1\% in the training data were used for prediction (about 400,000 SNPs; depending on the disease and the split). 

The fourth data set comes from a genetically heterogeneous stock of mice, consisting of 1904 individuals from 85 families, all descended from eight inbred progenitor strains \cite{Valdar:2006}. We used this data set to assess phenotype prediction performance of several methods. Multiple phenotype measurements are available for the data set, and we selected three phenotypes among them: percentage of CD8+ cells (CD8, $n=1410$),  mean corpuscular hemoglobin (MCH, $n=1580$) and body mass index (BMI, $n=1828$). We selected these phenotypes because they were previously used for comparing prediction performance of various methods \cite{Lee:2008, Legarra:2008, Campos:2009}, and they represent a wide range of narrow sense heritability: CD8 has a high heritability, MCH has a median heritability and BMI has a low heritability \cite{Valdar:2006b}. All phenotypes were already corrected for sex, age, body weight, season and year effects \cite{Valdar:2006}, and we further quantile normalized the phenotypes to a standard normal distribution. A total of 12,226 autosomal SNPs were available for all mice. For individuals with missing genotypes, we imputed missing values by the mean genotype of that SNP in their family. All polymorphic SNPs with minor allele frequency above 1\% in the training data were used for prediction (about 10,000 SNPs; depending on the phenotype and the split).

\subsection*{Simulations}

We used genotypes from the human height data set \cite{Yang:2010} 
described above and simulated phenotypes from the simple linear model (\ref{eqn:simple-linear})
with different assumptions for the distribution of effect sizes $\bbeta$. We consider two simulation scenarios where the true PVE is known and we simulated 20 independent sets of phenotype data in each case.

Scenario I: the effect sizes of causal SNPs come from a normal distribution. We randomly chose a fixed number of causal SNPs (10, 100, 1000, 10000) and simulated their effect sizes from a $\mbox{N}(0,1)$ distribution. We drew the errors from a normal distribution with variance chosen to achieve a given PVE (0.2 and 0.6). 

Scenario II: the effect sizes of causal SNPs come from a mixture of two normal distributions, such that a small group of causal SNPs have additional effects. We first randomly chose a large number of causal SNPs (10000), and among them, we further selected a small number of medium effect size SNPs (10 or 100) and used what were left as small effect size SNPs (9990 or 9900). We simulated the small effect sizes for all causal SNPs (10000) from a $\mbox{N}(0,1)$ distribution. Afterwards, we drew additional effect sizes (in addition to the small effects already drawn) for those medium effect SNPs (10 or 100) from a $\mbox{N}(0,1)$ distribution, and scaled these additional effect sizes further so that together they explained a fixed proportion of genetic variance, or PGE (0.1 and 0.2, for 10 and 100 medium effect size SNPs, respectively). Once we obtained the final effect sizes for all causal SNPs, we drew errors to achieve a given PVE (0.2 and 0.6).

\subsection*{Assessing Prediction Performance in Simulations}   
\label{sec:rpg}
\subsubsection*{MSPE and RPG}
We assess prediction accuracy mainly using mean square prediction error (MSPE), and a rescaled version of MSPE called relative predictive gain (RPG). The MSPE for predicting a future observation
in the simple linear model (\ref{eqn:simple-linear}) depends on comparing an estimated value of 
$\bbeta$, $\hat{\bbeta}$, the true value of $\bbeta$, and the error variance $\tau$, as follows:
\begin{align}
\mbox{MSPE}(\hat\bbeta; \bbeta, \tau)&:=E(\bx^T\hat \bbeta-y)^2\\
&=E( (\sum_{i=1}^p x_{i} (\hat\beta_i-\beta_i))^2)+\tau^{-1} \\
&=\sum_{i=1}^p \sum_{j=1}^p r_{ij} (\hat\beta_i-\beta_i)(\hat\beta_j-\beta_j)+\tau^{-1}\\
&\approx \sum_{|i-j|\leq 20} s_{ij}(\hat\beta_i-\beta_i)(\hat\beta_j-\beta_j)+\tau^{-1},
\end{align}
where $y$ is the phenotype for a future observation, $\bx$ is the corresponding $p$-vector of genotypes, $r_{ij}=E(x_ix_j)$ is the covariance between SNP $i$ and $j$. In practice, we approximate $r_{ij}$ with the sample covariance $s_{ij}=\frac{1}{n}\sum_{k=1}^n x_{ik}x_{jk}$, and we only consider $s_{ij}$ for neighboring SNPs that satisfy $|i-j|\leq 20$. This is because linkage dis-equilibrium (LD) decays with distance and remote SNPs are approximately independent with each other. The above definition of MSPE extends the definition in \cite{Guan:2011} to take into account correlations among neighboring SNPs.

We denote MSPE$_0$ as the MSPE obtained using only the mean of the phenotype (i.e. $\bar \by$) for prediction, and we define RPG as the rescaled version of MSPE following \cite{Guan:2011}:
\begin{equation}
\mbox{RPG}(\bbeta, \hat\bbeta):=\frac{\mbox{MSPE}_0-\mbox{MSPE}(\hat \bbeta; \bbeta, \tau)}{\mbox{MSPE}_0-\mbox{MSPE}(\bbeta; \bbeta, \tau)}.
\end{equation}

When applying these formulae for LMM or BSLMM note that 
we use estimates $\hat\bbeta$ for $\bbeta$ in (\ref{eqn:simple-linear}),
and not for $\tbbeta$ in (\ref{eqn_bslmm}). This ensures that the resulting
predictions take account of both the sparse effects $\bbeta$ and the random effect $\bu$ in (\ref{eqn_bslmm}). The way we obtain these estimates $\hat\bbeta$ is detailed below.

\subsubsection*{Correlation}
We also assess prediction accuracy using correlation. The correlation between the predicted value and the true value for a future observation in the simple linear model (\ref{eqn:simple-linear}) depends on $\hat{\bbeta}$, $\bbeta$ and $\tau$ as follows:
\begin{align}
\mbox{Cor}(\hat\bbeta; \bbeta, \tau)&:=\mbox{Cor}(\bx^T\hat \bbeta, y)\\
&=\frac{E((\sum_{i=1}^p x_i\hat\beta_i)(\sum_{j=1}^p x_j\beta_j))}{\sqrt{E((\sum_{i=1}^p x_i\hat\beta_i)^2) E((\sum_{j=1}^p x_j\beta_j)^2) /\mbox{PVE}(\beta, \tau) } }\\
&=\frac{(\sum_{i=1}^p\sum_{j=1}^p r_{ij}\hat\beta_i \beta_j)\sqrt{\mbox{PVE}(\beta, \tau)}}{\sqrt{(\sum_{i=1}^p\sum_{j=1}^p r_{ij}\hat\beta_i \hat\beta_j)(\sum_{i=1}^p\sum_{j=1}^p r_{ij}\beta_i \beta_j)}}\\
&\approx \frac{(\sum_{|i-j|\leq 20} s_{ij}\hat\beta_i \beta_j)\sqrt{\mbox{PVE}(\beta, \tau)}}{\sqrt{(\sum_{|i-j|\leq 20} s_{ij}\hat\beta_i \hat\beta_j)(\sum_{|i-j|\leq 20} s_{ij}\beta_i \beta_j)}}.
\end{align}

Again, when applying these formulae for LMM or BSLMM note that 
we use estimates $\hat\bbeta$ for $\bbeta$ in (\ref{eqn:simple-linear}),
and not for $\tbbeta$ in (\ref{eqn_bslmm}).

\subsection*{Details of BSLMM} 

\subsubsection*{Centering $\bX$ and $\bK$}
\label{sec:centering}
We assume that the genotypes in $\bX$ have been measured on bi-allelic markers, and that the genotypes at each marker are coded as 0, 1 or 2 copies of some reference allele. (For imputed genotypes we use the posterior mean genotype \cite{Guan:2008}.)
It occasionally simplifies the algebra to assume that each column of $\bX$ is centered to have mean 0; since the results will be the same with or without centering, we make this assumption throughout. It is also common to standardize the columns of $\bX$ to have unit variance. In contrast to centering, standardizing the columns will affect the results, and we do not standardize the columns in our applications here, although all our methods could be applied with the matrix standardized in this way. (Standardizing the columns of $\bX$ corresponds to making an assumption that rarer variants tend to have larger effects than common variants, and precisely that marker effect sizes tend to decay with the inverse of the genotype variance; see \cite{Wakefield:2009, Stephens:2009} for relevant discussion.) In summary, throughout this paper $X_{ij} = (x_{ij} - \bar{x}_j)$ where
$x_{ij}$ is the number of copies of the reference allele at marker $j$ in individual $i$, and $\bar{x}_j:= (1/n) \sum_i x_{ij}$.

To facilitate prior specification, in addition to centering the genotype matrix $\bX$, we also assume that the relatedness matrix $\bK$ is ``centered", in the sense that the random effects have mean zero: $\sum_{i=1}^n u_i=0$. This holds automatically for  $\bK \propto \bX \bX^T$, with $\bX$ centered. More generally it can be achieved by multiplying the relatedness matrix with a projection matrix on both sides: $\bM\bK\bM$, where $\bM=\bI_n-\mathbf 1_n(\mathbf 1_n^T\mathbf 1_n)^{-1}\mathbf 1_n^T$. The resulting transformed relatedness matrix is positive-semidefinite as long as the original relatedness matrix is positive-semidefinite.

\subsubsection*{Definition and derivation of expressions for $h$ and $\rho$}
We define $h,\rho$
\begin{align}
h(\pi, \sigma_a,\sigma_b)&:=\frac{E(\mbox{V}(\bX\tbbeta+\bu))}{E(\mbox{V}(\bX\tbbeta+\bu))+\tau^{-1}},\\
\rho(\pi,\sigma_a,\sigma_b)&:=\frac{E(\mbox{V}(\bX\tbbeta))}{E(\mbox{V}(\bX\tbbeta+\bu))},
\end{align}
where the function $V(\bx)$ is defined in equation \ref{Vfunc_def}, and the expectations are taken with respect to $(\tbbeta,\bu)$, conditional on hyper parameters ($\sigma_a,\sigma_b,\pi, \tau$). These conditional expectations are extensions of, and slight simplifications of, the similar expression for $h$ in \cite{Guan:2011}; the simplification
comes from taking expectations conditional on $\pi$ instead of conditional on $\bgamma$.
These definitions of $h$ and $\rho$ are motivated by approximating the expectations of PVE and PGE by the ratios of the expectations of the numerator and denominator. Both $h$ and $\rho$ take values between 0 and 1 and serve as rough guides to the expectations of PVE and PGE, respectively. 

The expectations in the above expressions, conditional on hyper-parameters ($\sigma_a^2, \sigma_b^2, \pi, \tau$), can be obtained as:
\begin{align}
\mbox{E}(\mbox{V}(\bX\tbbeta)|\sigma_a^2, \pi, \tau^{-1})&=\mbox{E}(\sum_{i=1}^p \mbox{V}(\bx_i\tbeta_i) |\sigma_a^2, \pi, \tau^{-1} )=p\pi s_a\sigma_a^2\tau^{-1}, \\
\mbox{E}(\mbox{V}(\bX\tbbeta+\bu)|\sigma_a^2, \sigma_b^2, \pi, \tau^{-1})&=\mbox{E}(\sum_{i=1}^p \mbox{V}(\bx_i\tbeta_i)+\mbox{V}(\bu)|\sigma_a^2, \sigma_b^2, \pi, \tau^{-1})=p\pi s_a\sigma_a^2\tau^{-1}+s_b\sigma_b^2\tau^{-1},
\end{align}
where $\bx_i$ is the $i$th column of $\bX$, $s_a=\frac{1}{np}\sum_{i=1}^p\sum_{j=1}^n x_{ij}^2$, $s_b=\frac{1}{n}\sum_{i=1}^n k_{ii}$, $x_{ij}$ and $k_{ij}$ are the $ij$th elements of matrices $\bX$ and $\bK$, respectively. The above derivation assumes centered genotypes and relatedness matrix. 

Plugging these approximations into the expressions (\ref{PVE_def}) and (\ref{PGE_def}) gives (\ref{h_def}) and (\ref{rho_def}).

\subsubsection*{Induced Priors on $\sigma_a^2$ and $\sigma_b^2$}

Solving (\ref{h_def}) and (\ref{rho_def}) for $\sigma_a^2$ and $\sigma_b^2$ as functions of $h$, $\rho$ and $\pi$
gives:
\begin{align}
\sigma_a^2&=\frac{h\rho}{(1-h)p\pi s_a}, \\
\sigma_b^2&=\frac{h(1-\rho)}{(1-h)s_b}.
\end{align}
The independent priors (\ref{h_prior}), (\ref{rho_prior}), (\ref{pi_prior}) for $(h,\rho,\pi)$ induce a joint prior on $(\sigma_a,\sigma_b, \pi)$: 
\begin{equation}
p(\sigma_a^2, \sigma_b^2, \pi )\propto \frac{p s_as_b}{(p\pi s_a \sigma_a^2+s_b\sigma_b^2+1)^2(p\pi s_a \sigma_a^2+s_b\sigma_b^2)},
\end{equation}
which is heavy tailed for ($\sigma_a^2$, $\sigma_b^2$) marginally (i.e. tail has a polynomial decay), a feature desirable in association studies \cite{Stephens:2009}. The priors also
have another nice property that the marker effect size variance $\sigma_a^2$ tends to decrease as the proportion of markers with an effect ($\pi$) increases.

\subsection*{PVE Estimation with LMM, BVSR and BSLMM}
\label{subsec:pveest}
We could estimate PVE using the posterior mean of the MCMC samples for all three models, and we do so for both BVSR and BSLMM. For LMM, we follow previous studies \cite{Yang:2010} and use an approximation to the PVE. In particular, we consider the LMM defined by equations (\ref{eqn_bslmm})-(\ref{eqn_uterm}) with $\tbbeta=0$, and we estimate PVE by the ratio of expectations
\begin{equation}
\hat{\mbox{PVE}}=\frac{ s_b \hat\sigma_b^2}{ s_b\hat\sigma_b^2+1}, 
\end{equation}
where $\hat\sigma_b^2$ is the REML estimate for the variance component. This formula can be viewed as a generalization of the form used in \cite{Yang:2010}, and is valid for any choice of centered relatedness matrix $\bK$.

% suitable for cases where the relatedness matrix have diagonal elements apart from one. 

To obtain the standard error of the above estimate, we compute the second derivative of the log restricted likelihood function with respect to $\sigma_b^2$ \cite{Zhou:2012} and evaluate it at $\hat \sigma_b^2$:
\begin{equation}
l_r''(\hat \sigma_b^2)=\frac{\partial l_r^2}{\partial^2 \sigma_b^2}|_{\sigma_b^2=\hat\sigma_b^2}=\frac{1}{2}\mbox{trace}(\bP\bK\bP\bK)-\frac{n-2}{2}\frac{2(\by^T\bP\by)(\by^T\bP\by)-(\by^T\bP\by)^2}{\by^T\bP\by}|_{\sigma_b^2=\hat\sigma_b^2},
\end{equation}
where $l_r$ denotes the log restricted likelihood, $\bH=\sigma_b^2 \bK+\bI_n$ and $\bP=\bH^{-1}-\bH^{-1}\mathbf 1_n(\mathbf 1_n\bH^{-1}\mathbf 1_n^T)^{-1}\mathbf 1_n^T\bH^{-1}$. Despite its complicated form, the second derivative can be easily and efficiently evaluated using recursions in \cite{Zhou:2012}. As the variance of $\hat \sigma_b^2$ is asymptotically $\mbox{v}(\hat\sigma_b^2)=-1/l''(\hat\sigma_b^2)$, using the delta method, we approximate the standard error of $\hat{\mbox{PVE}}$ by 
\begin{equation}
\mbox{se}(\hat{\mbox{PVE}})\approx \frac{s_b}{(s_b\hat\sigma_b^2+1)^2} \sqrt{\mbox{v}(\hat\sigma_b^2)}.
\end{equation}

As a check on correctness, we also used the posterior mean for PVE obtained from LMM-Bayes to estimate PVE. This gave almost identical results in all cases considered here and therefore only  results from LMM are presented.

\subsection*{Phenotype Prediction with LMM, BVSR and BSLMM}
\label{subsec:phenopred}

\subsubsection*{Real Data}

In the real data, we can perform prediction by estimating both the sparse effects ($\mu,\tbbeta$) and the random effects ($\bu$) in (\ref{eqn_bslmm}) from the training set, and use these to predict phenotypes in the test set. 

Let $\hat\mu_o$, $\hat\tbbeta_o$ and $\hat\bu_o$ denote estimates for the sparse and random effects 
obtained from the observed (training) sample. For BSLMM and BVSR these estimates are the posterior means for these parameters, estimated from MCMC samples (for BVSR $\hat\bu_o \equiv 0$). For LMM $\hat\bu_o$ is obtained as the conditional posterior mean of $\bu_o$ given the REML estimate for $\sigma_b^2$ (i.e.~BLUP). 
 
We then obtain predictions for a future (test) sample as follows.
For a general relatedness matrix $\bK$, we assume that the random effects for the observed and future samples follow a multivariate normal distribution
\begin{equation}
\begin{pmatrix} \bu_o \\ \bu_f \end{pmatrix} \sim \mbox{MVN}_{n_o+n_f}(\begin{pmatrix} 0 \\ 0 \end{pmatrix}, \begin{pmatrix} \bK_{oo} & \bK_{of} \\ \bK_{fo} & \bK_{ff} \end{pmatrix} ),
\end{equation}
where $n_o$ and $n_f$ are the sample size for observed (training) and future (test) data, respectively. Standard multivariate normal theory gives the conditional distribution 
\begin{equation}
\bu_f |\bu_o \sim \mbox{MVN}_{n_f} (\bK_{fo}\bK_{oo}^{-1}\bu_o, \bK_{ff}-\bK_{fo}\bK_{oo}^{-1}\bK_{of}).
\end{equation}

We use the conditional mean as an estimate for $\bu_f$ and thus the predicted phenotypes for future observations are
\begin{equation}
\hat \by_f=\mathbf 1_{n_f} \hat\mu_o+\bX_f\hat\tbbeta_o+\bK_{fo}\bK_{oo}^{-1}\hat\bu_o.
\end{equation}

\subsubsection*{Simulated Data}
\label{sec:predict_sim_detail}

In the simulated data we use RPG and correlation to assess prediction accuracy. To compute RPG and correlation we need to obtain
estimates for $\bbeta$ in the simple linear model (\ref{eqn:simple-linear}).
To do this, in the special case when the relatedness matrix $\bK=\frac{1}{p}\bX\bX^T$, we rewrite the
model (\ref{eqn_bslmm})  as
\begin{align}
\label{eqn_bslmm2}
\by&=\mathbf 1_n \mu+\bX\tbbeta+\bX \balpha+\bepsilon,   \\  
\bepsilon &\sim \mbox{MVN}_n(0, \tau^{-1} \bI_n),\\
\tbeta_i &\sim \pi \mbox{N}(0, \sigma_a^2\tau^{-1})+(1-\pi)\delta_0, \\
\alpha_i &\sim  \mbox{N}(0, \sigma_b^2/(p\tau)), \label{eqn:alpha}
\end{align}
where we can think of the $p$-vector $\balpha$ as representing the ``small" effect sizes present at every locus.
The special case of $\pi=0$ ($\tbbeta \equiv 0$) gives LMM, and the special case of $\sigma_b^2=0$ ($\balpha \equiv 0$) gives BVSR.
Note that $\balpha+\tbbeta= \bbeta$, so summing estimates of $\balpha$ and $\tbbeta$ yields an estimate for $\bbeta$ in (\ref{eqn:simple-linear}).  

For LMM we estimate $\balpha$ by its conditional expectation
\begin{equation}
\hat\balpha=\frac{\hat\sigma_b^2}{p}\bX^T(\hat \sigma_b^2\bK+\bI_{n})^{-1}\by,
\end{equation}
where $\hat\sigma_b^2$ is the REML estimate of the variance component in the observed data. Since $\tbbeta \equiv 0$ in LMM, this estimate
for $\balpha$ provides the required estimate for $\bbeta$ in (\ref{eqn:simple-linear}).

For BVSR, we use the posterior mean of $\tbbeta$ (since $\balpha \equiv 0$).

For BSLMM, we use Rao-Blackwellisation to obtain
an approximation to the posterior mean for $\balpha$ (Text S2), and then add this to the (approximate) posterior mean for $\tbbeta$ obtained from the MCMC sampler
to obtain an approximation for the posterior mean of $\bbeta$ in (\ref{eqn:simple-linear}).

\subsection*{Other Methods}

\begin{enumerate}
\item { \bf LMM:} We fit LMM using the GEMMA algorithm \cite{Zhou:2012}. 
\item {\bf BVSR:} We fit BVSR by fixing $\rho=1$ in BSLMM using our software. This gives slightly better results, and is faster than the BVSR software piMASS (version 0.90) \cite{Guan:2011}, in all examples considered here.
\item {\bf LMM-Bayes:}  We fit this by fixing $\rho=0$ in BSLMM using our software. 
\item {\bf Bayesian Lasso:} This \cite{Park:2008} assumes a double-exponential prior for each coefficient $\beta_i$  in (\ref{eqn:simple-linear}):
\begin{equation}
\beta_i | \lambda\sim \mbox{DE}(0, \lambda^{-1}), \quad \lambda^2 \sim \mbox{Gamma}(\kappa_1, \kappa_2), \quad \tau^{-1}\sim \mbox{IG}(\kappa_3, \kappa_4),
\end{equation}
where DE denotes the double exponential (Laplace) distribution with mean 0 and scale parameter $\lambda^{-1}$, and Gamma denotes a Gamma distribution with shape and rate parameters. We set $\kappa_1=0.55, \kappa_2=10^{-6}, \kappa_3=1/2$ and  $\kappa_4=1/2$ following previous studies \cite {Campos:2009, Makowsky:2011}. We use a conjugate Gamma prior for $\lambda^2$ as in \cite{Park:2008} instead of a Beta prior for $\lambda/100$ as in \cite{Campos:2009}. We used the R package BLR \cite{Campos:2009} to sample from the posterior distribution of $\bbeta$. 
\item {\bf BayesA-Flex:} This assumes a scaled t-distribution for each coefficient $\beta_i$ in (\ref{eqn:simple-linear}):
\begin{equation}
\beta_i | \sigma \sim t(0, \nu, \sigma^2), \quad \sigma^2 \sim \mbox{IG}(\kappa_1, \kappa_2), \quad \tau^{-1}\sim \mbox{IG}(\kappa_3, \kappa_4),
\end{equation}
where IG stands for the inverse gamma distribution. Following previous studies, we set the degree of freedom parameter $\nu$ to 4 \cite{Meuwissen:2001, Hayes:2010, Habier:2011} and set $\kappa_3=1/2$ and  $\kappa_4=1/2$ \cite {Campos:2009}. We also consider the posterior distribution where $\kappa_1\to 0$ and $\kappa_2\to 0$. The above model is similar to BayesA \cite{Meuwissen:2001}, but with a key difference in the way the scaling parameter $\sigma^2$ is treated: BayesA fixes $\sigma^2$ to some pre-specified value, whereas here we specify a prior for $\sigma^2$ and allow it to be estimated from the data (and hence the name ``BayesA-Flex").  Using BayesA in this data set gives poor results \cite{Legarra:2008} (and data not shown); but estimating the scaling parameter greatly improves prediction performance. We modified the R package BLR \cite{Campos:2009} to obtain posterior samples from this model. The modified code is freely available online.
\item {\bf BayesC$\pi$:} we fit this using the online software GenSel \cite{Habier:2011}.  
\item {\bf BSLMM-EB:} we fit this using our BSLMM software, fixing $\sigma_b^2$ to the REML estimate $\hat\sigma_b^2$ from the null model (i.e. LMM). This approximation avoids updating $\sigma_b^2$ in each iteration of the MCMC, and is one of the several approximation strategies used by previous studies to alleviate the computation burden of models similar to BSLMM \cite{Lee:2008, Segura:2012}. Intuitively, by fixing the variance component to its null estimate, BSLMM-EB discourages the inclusion of large effect SNPs into the sparse effects term and risks underestimating their effect sizes. Therefore, this approximation may reduce the prediction performance of BSLMM, especially when there are large effect SNPs. We confirm this in the real data set.
\end{enumerate}
For all MCMC based methods except for BayesC$\pi$, we run 2.1 million iterations with the first 0.1 million iterations as burn-in steps. For BayesC$\pi$, due to web server restriction, we run 1.1 million iterations with the first 0.1 million iterations as burn-in.

\clearpage
\newpage

\section*{Text S2 Detailed MCMC Strategy for BSLMM}
To simplify notation, we assume in this section that $\by$ is centered. We use Markov chain Monte Carlo to obtain posterior samples of ($h, \rho, \pi, \bgamma$) on the product space $(0,1)\times(0,1)\times(0,1)\times\{0,1\}^p$, which is given by
\begin{equation}
P(h,\rho, \pi, \bgamma |\by)\propto P(\by|h, \rho, \pi, \bgamma)P(h)P(\rho)P(\bgamma|\pi)P(\pi).
\end{equation}

In the above equation, we explored the fact that the parameters $\tbbeta$, $\bu$ and $\tau$ can be integrated out analytically to compute the marginal likelihood $P(\by|h, \rho, \pi, \bgamma)$. The marginal likelihood is
\begin{equation}
P(\by|h, \rho, \pi, \bgamma)\propto |\bH|^{-\frac{1}{2}} |\sigma_a^{-2}\bOmega|^{\frac{1}{2}} (\by^T\bP \by)^{-\frac{n}{2} },
\end{equation}
where $\bH(\sigma_b^2)=\sigma_b^2 \bK+\bI_n$, $\bOmega(\sigma_a^2, \sigma_b^2, \gamma)=(\bX_{\gamma}^T\bH^{-1}\bX_{\gamma}+\sigma_a^{-2}\bI_{|\gamma|})^{-1}$, $\bP(\sigma_a^2, \sigma_b^2, \gamma)=\bH^{-1}-\bH^{-1}\bX_{\gamma}\bOmega \bX_{\gamma}^T\bH^{-1}$. Notice again that $\sigma_a^2$ is a function of $h, \rho$ and $\pi$, while $\sigma_b^2$ is a function of $h$ and $\rho$.

To efficiently evaluate the marginal likelihood, we perform an eigen decomposition of the relatedness matrix $\bK=\bU\bD\bU^T$  at the beginning of the MCMC, where $\bU$ is the matrix of eigen vectors and $\bD$ is a diagonal matrix of eigen values. We transform both the phenotype vector and the genotype matrix by multiplying the eigen matrix and calculate $\bU^T\by$ and $\bU^T\bX$. Afterwards, as has been shown previously, the calculations of the determinant and the inverse of matrix $\bH$, as well as the vector-matrix-vector form $\by^T\bP \by$, in each iteration of the MCMC, are easy to perform \cite{Lippert:2011, Zhou:2012}. 

We use a standard Metropolis-Hastings algorithm to draw posterior samples of the hyper-parameters $(h, \rho, \pi, \bgamma)$ based on the above marginal likelihood. Following \cite{Guan:2011}, we use a rank based proposal distribution for $\bgamma$, and use random walk proposals based on uniform distributions for $h$, $\rho$ and $\log(\pi)$. In particular, we first obtain single-SNP $p$ values using a standard LMM with GEMMA algorithm \cite{Zhou:2012}, and then rank SNPs based on these $p$ values from small to large. Our aim is to use a proposal distribution for $\bgamma$ that puts more weights on SNPs that are ranked higher by the single SNP tests, and to do this we consider a mixture distribution $Q_p=0.3U_p+0.7G_p$, where $U_p$ is a uniform distribution on $1, \cdots, p$ and $G_p$ is a geometric distribution truncated to $1, \cdots, p$ with its parameter chosen to give a mean of 2000. We denote $\bgamma^+=\{i: \gamma_i=1\}$ and we propose the new $\bgamma$ by randomly choose one of the following steps:
\begin{itemize}
\item add a covariate with probability 0.4: generate $r$ from $Q_p$ until the covariate with rank $r$ is not in $\bgamma^+$, then add this covariate to $\bgamma^+$
\item remove a covariate with probability 0.4: pick a covariate in $\bgamma^+$ uniformly at random and remove it from $\bgamma^+$
\item switch a pair of covariates with probability 0.2: pick up two covariates by the above two steps, and switch their indicator values
\end{itemize}

For the other hyper-parameters, we update $\log(\pi)$ by adding a random variable from U(-0.05, 0.05) to the current value, and update $h$ and $\rho$ by adding a random variable from U(-0.1, 0.1) to the current values. New values of $h$ and $\rho$ that lie outside the boundary [0,1] are reflected back. 

In addition to the above local proposal distributions, we also use a ``small world proposal" which improves theoretical MCMC convergence \cite{Guan:2007}. In brief, with probability 0.33 in each iteration, we make a longer-range proposal by compounding many local moves, where the number of compounded local moves is draw uniformly from 1 to 20. 

For each sampled values of $(h, \rho, \pi, \bgamma)$, we further obtain samples of $\tau$ and $\tbbeta$ using the conditional distributions $P(\tau| \by, h, \rho, \pi, \bgamma)$ and $P(\tbbeta|\by, h, \rho, \pi, \bgamma, \tau) $ listed below:
\begin{align}
\tau|\by, h, \rho, \pi, \bgamma &\sim \mbox{Gamma}(\frac{n}{2}, \frac{\by^T\bP\by}{2}), \\
\tbbeta_{\bgamma}|\by, h, \rho, \pi, \bgamma, \tau &\sim \mbox{MVN}_{|\gamma|} (\bOmega \bX_{\gamma}^T\bH^{-1}\by, \tau^{-1}\bOmega), \\
\tbbeta_{-\bgamma}|\by, h, \rho, \pi, \bgamma, \tau &\sim \bdelta_0.
\end{align}
Afterwards, we sample $\bu$ based the conditional distribution $P(\bu|\by, h, \rho, \pi, \bgamma, \tau, \tbbeta)$:
\begin{align}
\bu|\by, h, \rho, \pi, \bgamma, \tau, \tbbeta \sim \mbox{MVN}_n (\sigma_b^2 \bK \bH^{-1}(\by-\bX_{\gamma} \tbbeta_{\gamma}), \sigma_b^2 \bK\bH^{-1}\tau^{-1}).
\end{align}
However, we do not sample $\bu$ directly from the above $n$-dimensional multivariate normal distribution. Instead, we sample $\bU^T\bu$ (and we never need to obtain $\bu$), as the conditional distribution of each element in $\bU^T\bu$ is a normal:
\begin{align}
\bU^T\bu|\by, h, \rho, \pi, \bgamma, \tau, \tbbeta \sim \mbox{MVN}_n (\sigma_b^2 \bD (\sigma_b^2\bD+\bI)^{-1}(\bU^T\by-\bU^T\bX_{\gamma} \tbbeta_{\gamma}), \sigma_b^2 \bD (\sigma_b^2\bD+\bI)^{-1}\tau^{-1}).
\end{align}
where the covariance matrix is diagonal. 

For each sampled value of ($\tbbeta$, $\bu$, $\tau$), we obtain samples of PVE and PGE based on equations (\ref{PVE_def}) and (\ref{PGE_def}). 

When required (e.g. for evaluating RPG in simulation studies), in the special case
$\bK=\bX\bX^T/p$, we also obtain the (approximate) posterior mean of $\balpha$ in the alternative model formulation (\ref{eqn_bslmm2})-(\ref{eqn:alpha}). This is achieved without sampling $\balpha$ in each iteration
using the fact that the full conditional distribution of $\balpha$ given other sampled values is
\begin{equation}
\balpha |\by, h, \rho, \pi, \bgamma, \tau, \tbbeta \sim \mbox{MVN}_n(\sigma_{b}^2 p^{-1}\bX^T\bH^{-1}(\by-\bX_{\bgamma}\tbbeta_{\bgamma}), \sigma_{b}^2(p^{-1}\bI_p-p^{-2}\sigma_{b}^{2}\bX^T\bH^{-1}\bX)\tau^{-1}),
\end{equation}
which leads to the Rao-Blackwellised approximation for the posterior mean of $\balpha$: 
\begin{equation}
\label{eqn_alpha}
\hat\balpha=\frac{1}{T}\sum_{t=1}^T\mbox{E}(\balpha |\by, h^{(t)}, \rho^{(t)}, \pi^{(t)}, \bgamma^{(t)}, \tau^{(t)}, \tbbeta^{(t)})=\frac{1}{p}\bX^T\frac{1}{T}\sum_{t=1}^T(\sigma_{b}^{(t)})^2 (\bH^{(t)})^{-1} (\by-\bX_{\bgamma^{(t)}} \tbbeta_{\bgamma^{(t)}}^{(t)}),
\end{equation}
where $T$ is the total number of MCMC iterations, and the superscript $(t)$ denotes the $t$th MCMC sample. Notice that we only need to do the $p$ dimensional matrix-vector multiplication once at the end. 

When $|\gamma|$ is large, the most time consuming part of our MCMC scheme for fitting BSLMM and BVSR is the calculation of $\Omega$. The per-iteration computation time of the above algorithm is comparable to that of BVSR \cite{Guan:2011} with linear complexity in the number of individuals but quadratic complexity in $|\bgamma|$. In practice, to reduce the computation burden, we set a maximal value for $|\bgamma|$ (300 for simulations and the two human data sets, 600 for the mouse data set). Setting the maximal value to a larger number (600) in simulations improves results only subtly, even for scenarios where a large number of causal SNPs is present.

\clearpage
\newpage

\section*{Text S3 The Probit BSLMM and Binary Traits}
\subsection*{MCMC strategy}

We use ``probit BSLMM" to refer to a BSLMM with a probit link to model binary traits:
\begin{equation}
P(y_i=1|\bx_i,\tbbeta, u_i)=1-P(y_i=0|\bx_i,\tbbeta, u_i)=\Phi(\mu+\bx_i\tbbeta+u_i) \quad (i=1,\cdots, n),
\end{equation}
where $y_i$ is the binary trait for $i$th individual, $\bx_i$ is the $i$th row-vector of $\bX$, $u_i$ is $i$th element of random effects vector $\bu$ and $\Phi$ is the cumulative distribution function (CDF) of the standard normal distribution. Following \cite{Albert:1993}, we introduce a vector of auxiliary variables $\bz$ and obtain the equivalent latent variable model as:
\begin{align}
y_i &=\left\{\begin{array}{l l}
    1 & \quad \text{if $z_i>0$}\\
    0 & \quad \text{if $z_i\leq 0$}\\
  \end{array} \right. , \\
z_i&=\mu+\bx_i\tbbeta+u_i+\epsilon_i \quad \epsilon_i\sim \mbox{N}(0, 1),
\end{align}
where $z_i$ is $i$th element of vector $\bz$. 

We use the same prior specifications for the hyper-parameters as described in the main text (except that $\tau=1$ here). We use a similar MCMC strategy as described in Text S2 to sample posteriors, with an additional step to sample the posteriors of the latent variables $\bz$ using the conditional distribution $P(\bz|\by, \bgamma, \tbbeta, \bu)$:
\begin{align}
z_i|y_i=1, \tbbeta, u_i &\sim    \mbox{N}(\mu+\bx_{\bgamma i}\tbbeta+u_i, 1) \quad \text{left truncated at 0},\\
z_i|y_i=0, \tbbeta, u_i &\sim    \mbox{N}(\mu+\bx_{\bgamma i}\tbbeta+u_i, 1) \quad \text{right truncated at 0}.
\end{align}
We denote $\bar z$ as the sample mean of $\bz$, or $\bar z=\frac{1}{n}\sum_{i=1}^n z_i$. Conditional on the latent variables $\bz$, the posterior sampling for the hyper-parameters ($h, \rho, \pi, \bgamma$) is based on the marginal likelihood $P(h,\rho, \pi, \bgamma |\bz)$, which is slightly different from that in Text S2 as we do not integrate out $\tau$ here:
\begin{equation}
P(\bz|h, \rho, \pi, \bgamma)\propto |\bH|^{-\frac{1}{2}} |\sigma_a^{-2}\bOmega|^{\frac{1}{2}} e^{-\frac{1}{2} (\bz-\mathbf 1_n \bar z)^T\bP (\bz-\mathbf 1_n \bar z)}.
\end{equation}
After obtaining the posterior samples of the hyper-parameters, we sample the posteriors of $\tbbeta$ and $\bu$ using conditional distributions identical to those listed in Text S2 by setting $\tau=1$. Finally, we sample $\mu$ based on the conditional distribution $P(\mu| \bz, \bgamma, \tbbeta, \bu)$:
\begin{equation}
\mu |\bz, \bgamma, \tbbeta, \bu \sim \mbox{N}(\frac{1}{n}\mathbf 1_n^T(\bz-\bX_{\bgamma}\tbbeta_{\bgamma}-\bu), \frac{1}{n}).
\end{equation}

For efficient calculation of the above marginal likelihood function $P(\bz|h, \rho, \pi, \bgamma)$, we use the same strategy as described in Text S2. However, as a transformation of the latent vector $\bz$ to $\bU^T\bz$ as well as transformations of $\bU^T\bu$ and $\bU^T\bX\tbbeta$ back to $\bu$ and $\bX\tbbeta$ are needed in every Gibbs iteration, the per-iteration computational cost of the probit BSLMM increases quadratically with the number of individuals. 

\subsection*{Application to Mouse Data}
To generate a binary data set on which to illustrate the probit BSLMM and compare its performance with BSLMM, we use the mouse data from the main
text, transforming the quantitative values of the three traits into binary values by assigning the half individuals with higher quantitative values to 1 and the other half to 0. We consider two different approaches here: (linear) BSLMM and probit BSLMM. The BSLMM can be viewed as a first order approximation to its probit counterpart. We use Brier score in the test sample to evaluate prediction performance. For BSLMM, we threshold the predicted probability values that are above 1 to be exact 1 and those below 0 to be exact 0. We contrast the performance of the probit BSLMM against BSLMM by calculating the Brier score difference, where a positive value indicates worse performance than BSLMM.

Figure \ref{fig:real_binary} shows Brier score differences for the three traits. Interestingly, for the three traits here, treating binary values as quantitative traits using BSLMM works better than modeling them directly using the probit BSLMM. This may partly reflect numerical inaccuracies due to the greater computational burden of fitting the probit BSLMM.

\subsection*{Correction factor for estimating PVE for case-control data}
Here, we provide an alternative way to derive the correction factor, that appeared in \cite{Lee:2011}, for transforming PVE estimate in the observed scale back to that in the liability scale. Our approach is based on Taylor series approximation. To simplify notation, we denote $k_p=P_{p}(y_i=1)$ as the case proportion in the population, $k_{a}=P_{a}(y_i=1)$ as the case proportion in the ascertained case-control sample, $\Phi$ as the normal CDF (cumulative distribution function), $\phi$ as the normal PDF (probability distribution function), $\mu_p$ satisfies $\Phi(\mu_p)=k_p$, $\mu_a$ satisfies $\Phi(\mu_a)=k_a$, and $z_p=\phi(\mu_p)$.

First, we assume, following \cite{Lee:2011}, a probit model on the population scale:
\begin{equation}
P_p(y_i=1| \bx_i, \tbbeta, u_i)=1-P_p(y_i=0| \bx_i, \tbbeta, u_i)=\Phi(\mu_p+\bx_i\tbbeta+u_i) \quad (i=1,\cdots, N),
\end{equation}
where $N$ is the population sample size. 

The conditional distribution in the ascertained case-control sample can be derived by Bayes theorem
\begin{equation}
P_a(y_i=1| \bx_i, \tbbeta, u_i)=\frac{P_a(\bx_i, u_i | y_i=1, \tbbeta)P_a(y_i=1 | \tbbeta)}{P_a(\bx_i, u_i | \tbbeta)} \quad (i=1,\cdots, n),
\end{equation}
where $n$ is the case-control sample size. 

We notice that $P_a(\bx_i, u_i | y_i=1, \tbbeta)=P_p(\bx_i, u_i | y_i=1, \tbbeta)$ holds for ideal case-control studies, as cases in the ascertained sample are selected randomly from all cases in the population. We assume further $P_p(y_i=1|\tbbeta)\approx P_p(y_i=1)=k_p$ and $P_a(y_i=1|\tbbeta)\approx P_a(y_i=1)=k_a$, that the probability of being a case does not depend on parameters, an assumption commonly made (see e.g. \cite{Imbens:1992}) and likely hold when parameters are close to their true values. We further denote a normalizing constant $Z=\frac{P_a(\bx_i, u_i|\tbbeta)}{P_p(\bx_i, u_i|\tbbeta)}$, and we have
\begin{equation}
P_a(y_i=1| \bx_i, \tbbeta, u_i)\approx \frac{1}{Z}\frac{k_a}{k_p}\Phi(\mu_p+\bx_i\tbbeta+u_i), 
\end{equation}
and similarly
\begin{equation}
P_a(y_i=0| \bx_i, \tbbeta, u_i)\approx \frac{1}{Z}\frac{1-k_a}{1-k_p}(1-\Phi(\mu_p+\bx_i\tbbeta+u_i)), 
\end{equation}
which give the normalizing constant 
\begin{equation}
Z=\frac{k_a}{k_p}\Phi(\mu_p+\bx_i\tbbeta+u_i)+\frac{1-k_a}{1-k_p}(1-\Phi(\mu_p+\bx_i\tbbeta+u_i)).
\end{equation}
We expand the above two likelihoods using Taylor series expansion with respect to $\bx_i\tbbeta+u_i$ at $0$. If we use the linear term only for approximation, we obtain
\begin{align}
P_a(y_i=1| \bx_i, \tbbeta, u_i)&\approx k_a+\frac{k_a(1-k_a)z_p}{k_p(1-k_p)}(\bx_i\tbbeta+u_i), \\
P_a(y_i=0| \bx_i, \tbbeta, u_i)&\approx 1-k_a-\frac{k_a(1-k_a)z_p}{k_p(1-k_p)}(\bx_i\tbbeta+u_i).
\end{align}
In other words, the expected value of individual binary label can be approximated by
\begin{align}
E(y_i)=P_a(y_i=1| \bx_i, \tbbeta, u_i)&\approx k_a+\frac{k_a(1-k_a)z_p}{k_p(1-k_p)}(\bx_i\tbbeta+u_i),
\end{align}
which suggests using a linear mixed model to treat binary values as quantitative traits to infer the parameters. The estimated PVE on the observed scaling using a linear mixed model is
\begin{align}
\hat {\mbox{PVE}_o}=(\frac{k_a(1-k_a)}{k_p(1-k_p)}z_p)^2\frac{V(\bx\hat{\tbbeta}+\hat u_i)}{V(\by)}=\frac{k_a(1-k_a)z_p^2}{k_p^2(1-k_p)^2}V(\bx\hat{\tbbeta}+\hat u_i)\approx \frac{k_a(1-k_a)z_p^2}{k_p^2(1-k_p)^2} \mbox{PVE}_l,
\end{align}
where $\hat {\mbox{PVE}_o}$ is the PVE estimate on the observed scale, and $\mbox{PVE}_l$ is the true PVE on the liability scale. 

Therefore, we can use the correction factor $\frac{k_p^2(1-k_p)^2}{k_a(1-k_a)z_p^2}$ to transform the PVE estimate on the observed scale back to that on the liability scale.

\clearpage
\newpage

\section*{Figures}

\clearpage
\newpage

\clearpage
\newpage

\clearpage
\newpage

\clearpage
\newpage

\setcounter{figure}{0}
\makeatletter 
\renewcommand{\thefigure}{S\@arabic\c@figure}

\clearpage
\newpage

\begin{figure}[!ht]
\centering
\includegraphics[width=8.3cm,keepaspectratio=true]{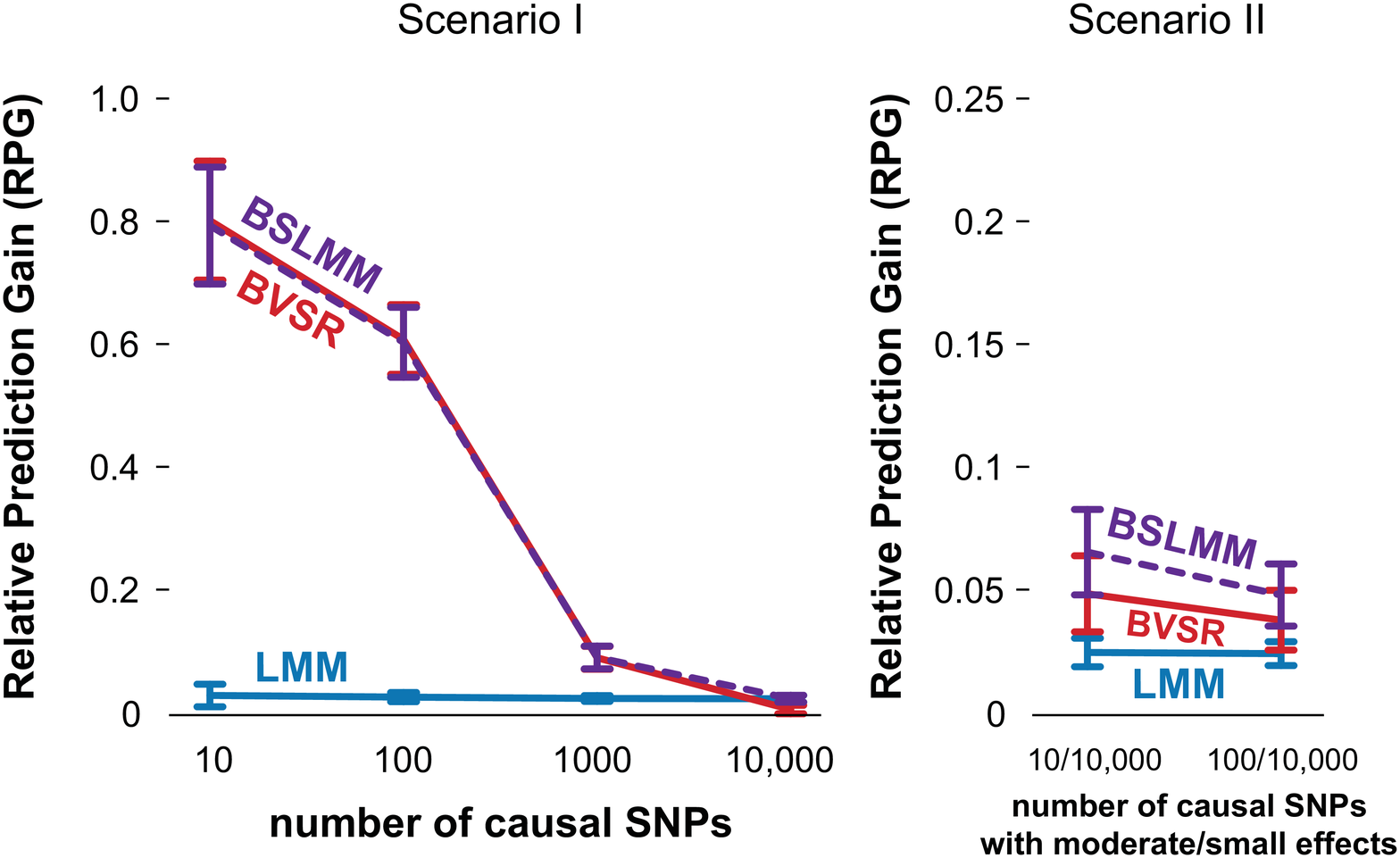}
\caption{Comparison of prediction performance of LMM (blue), BVSR (red) and BSLMM (purple) in two simulation scenarios, where all causal SNPs are excluded from the data. Performance is measured by Relative Predictive Gain (RPG). True PVE=0.6. Means and standard deviations (error bars) are based on 20 replicates. The x-axis show the number of causal SNPs (Scenario I) or the number of medium/small effect SNPs (Scenario II). }
\label{fig:sim_pred_exclude_rpg}
\end{figure}

\clearpage
\newpage

\begin{figure}[!ht]
\centering
\includegraphics[width=8.3cm,keepaspectratio=true]{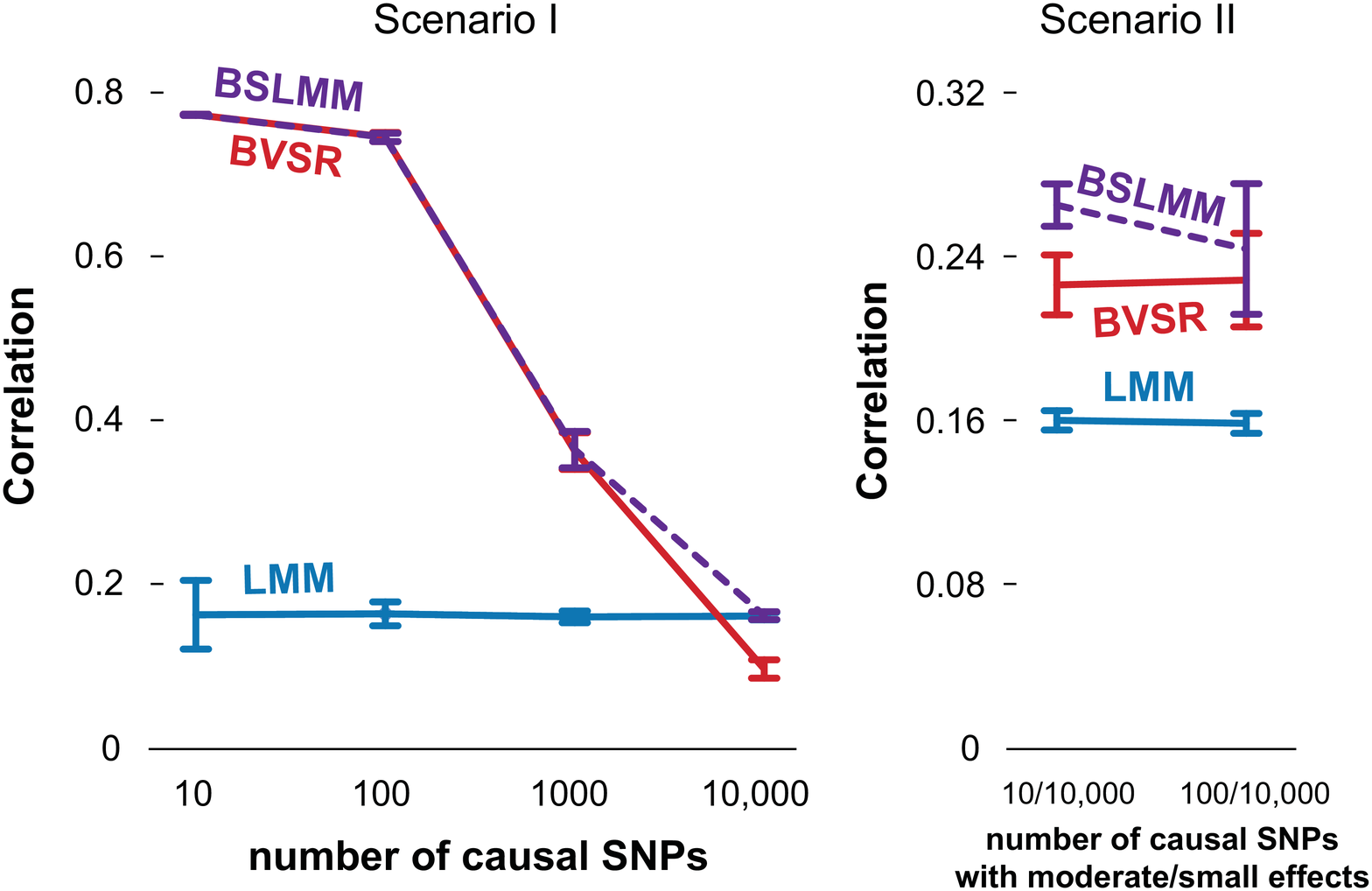}
\caption{Comparison of prediction performance of LMM (blue), BVSR (red) and BSLMM (purple) in two simulation scenarios, where all causal SNPs are included in the data. Performance is measured by correlation. True PVE=0.6. Means and standard deviations (error bars) are based on 20 replicates. The x-axis show the number of causal SNPs (Scenario I) or the number of medium/small effect SNPs (Scenario II). }
\label{fig:sim_pred_cor}
\end{figure}

\clearpage
\newpage

\begin{figure}[!ht]
\centering
\includegraphics[width=8.3cm,keepaspectratio=true]{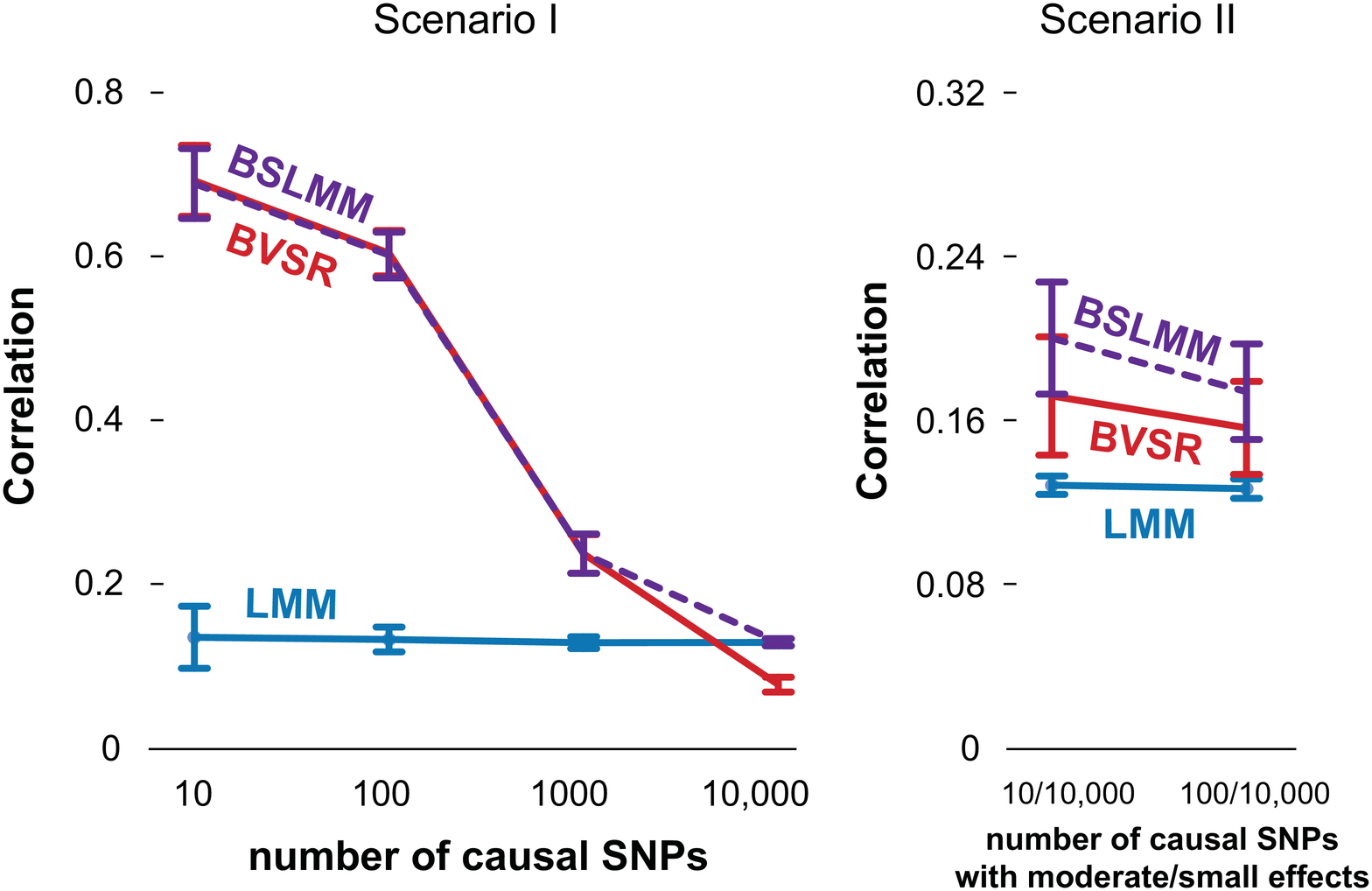}
\caption{Comparison of prediction performance of LMM (blue), BVSR (red) and BSLMM (purple) in two simulation scenarios, where all causal SNPs are excluded from the data. Performance is measured by correlation. True PVE=0.6. Means and standard deviations (error bars) are based on 20 replicates. The x-axis show the number of causal SNPs (Scenario I) or the number of medium/small effect SNPs (Scenario II). }
\label{fig:sim_pred_exclude_cor}
\end{figure}

\clearpage
\newpage

\begin{figure}[!ht]
\begin{center}
\includegraphics[width=17.35cm,keepaspectratio=true]{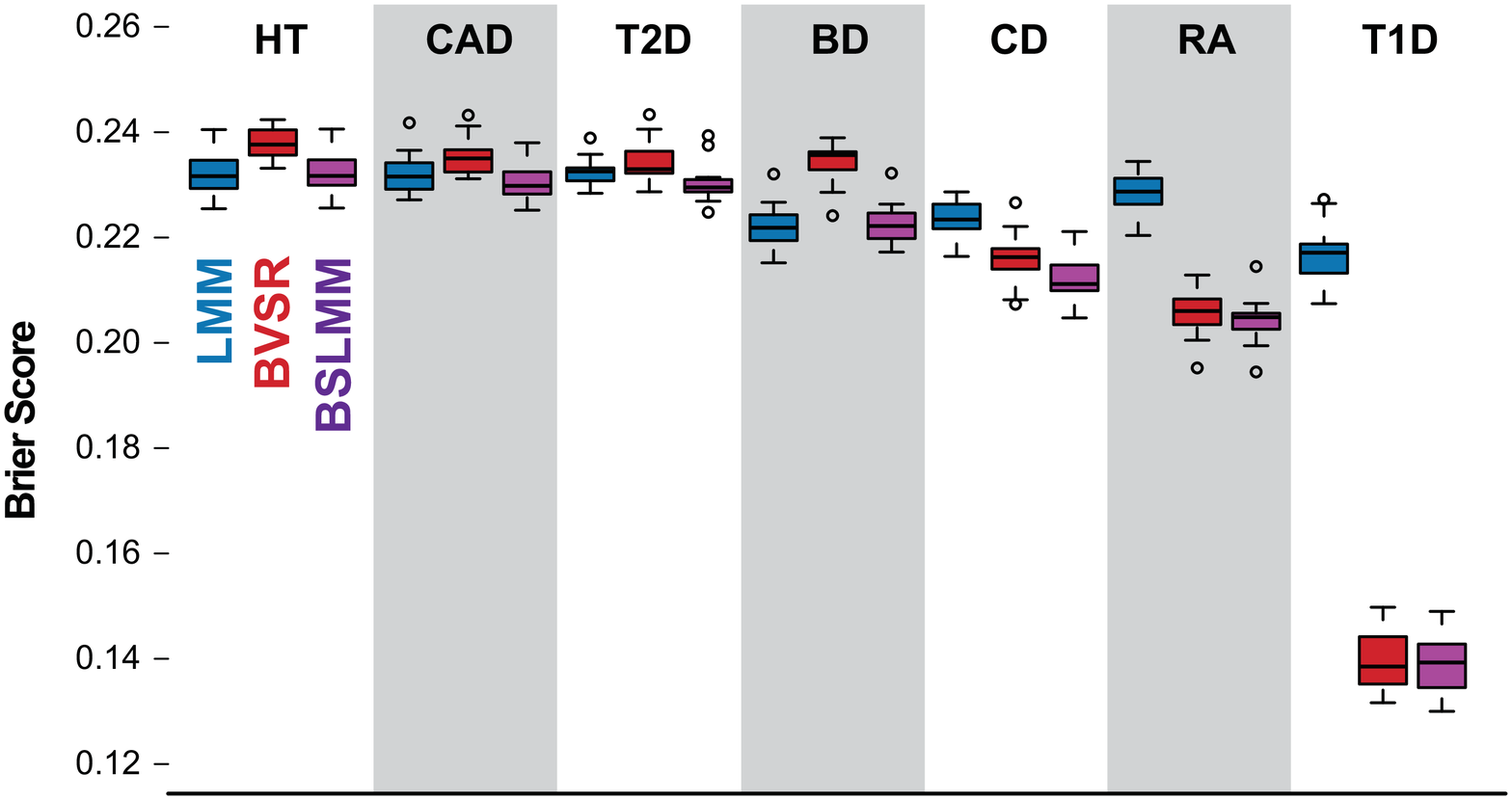}
\end{center}
\caption{Comparison of prediction performance of LMM (blue), BVSR (red) and BSLMM (purple) for seven diseases in the WTCCC data set. Performance is measured by Brier score, where a lower score indicates better performance. The order of the diseases is based on the performance of BSLMM. The mean and standard deviation of Brier scores for BSLMM in the seven diseases are 0.232 (0.004) for HT, 0.230 (0.004) for CAD, 0.230 (0.003) for T2D, 0.222 (0.004) for BD, 0.211 (0.004) for CD, 0.204 (0.004) for RA, 0.139 (0.006) for T1D.}
\label{fig:real_pred_wtccc_brier}
\end{figure}

\clearpage
\newpage

\begin{figure}[!ht]
\begin{center}
\includegraphics[width=17.35cm,keepaspectratio=true]{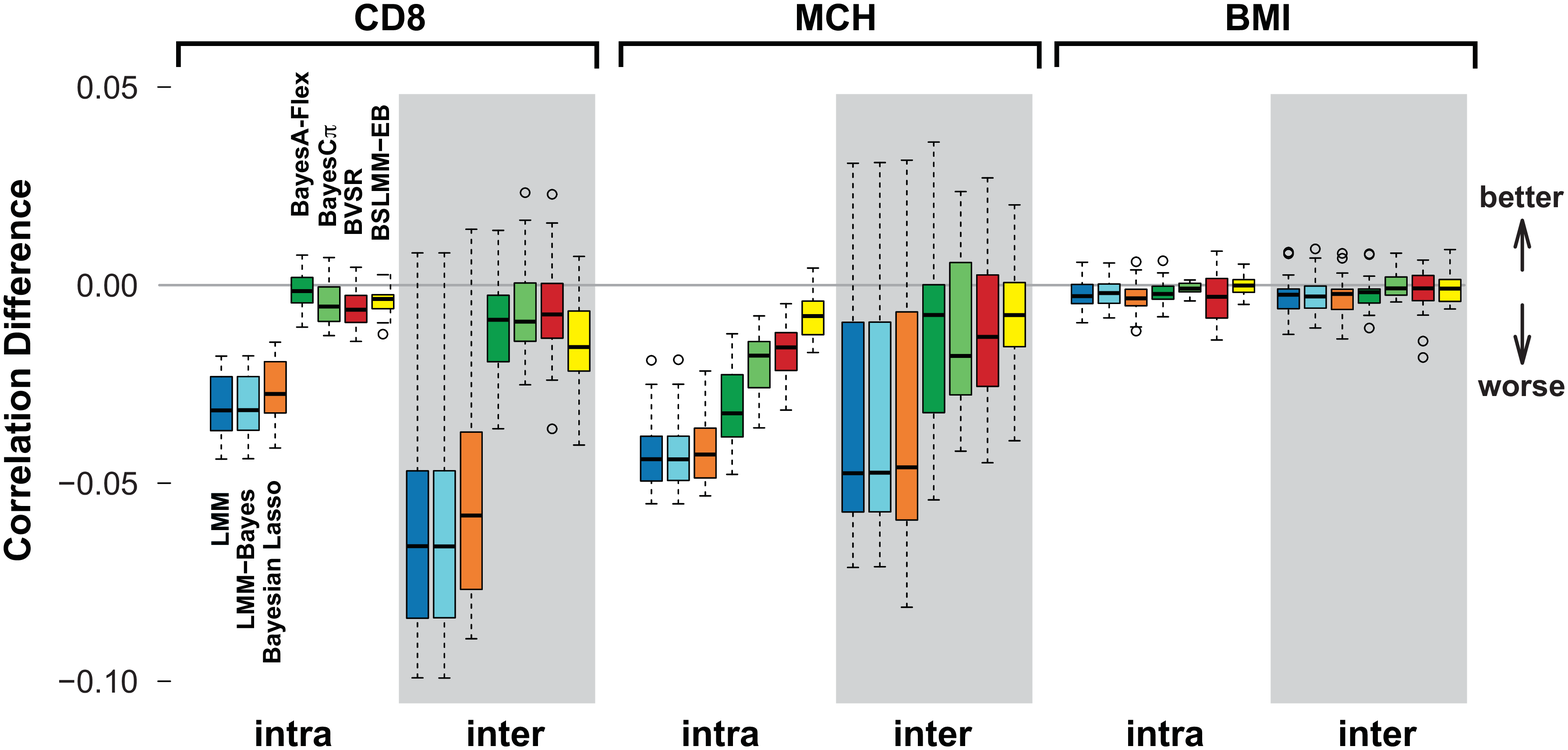}
\end{center}
\caption{Comparison of prediction performance of several models with BSLMM for three traits in the heterogenous stock mouse data set. Performance is measured by correlation difference with respect to BSLMM, where a positive value indicates better performance than BSLMM. The x-axis shows two different ways to split the data into a training set and a test set, each with 20 replicates. The mean correlation of BSLMM for the six cases are 0.72, 0.61, 0.61, 0.47, 0.21 and 0.14, respectively. }
\label{fig:real_pred_mouse_cor}
\end{figure}

\clearpage
\newpage

\begin{figure}[htb!]
\centering
\includegraphics[width=16cm,keepaspectratio=true]{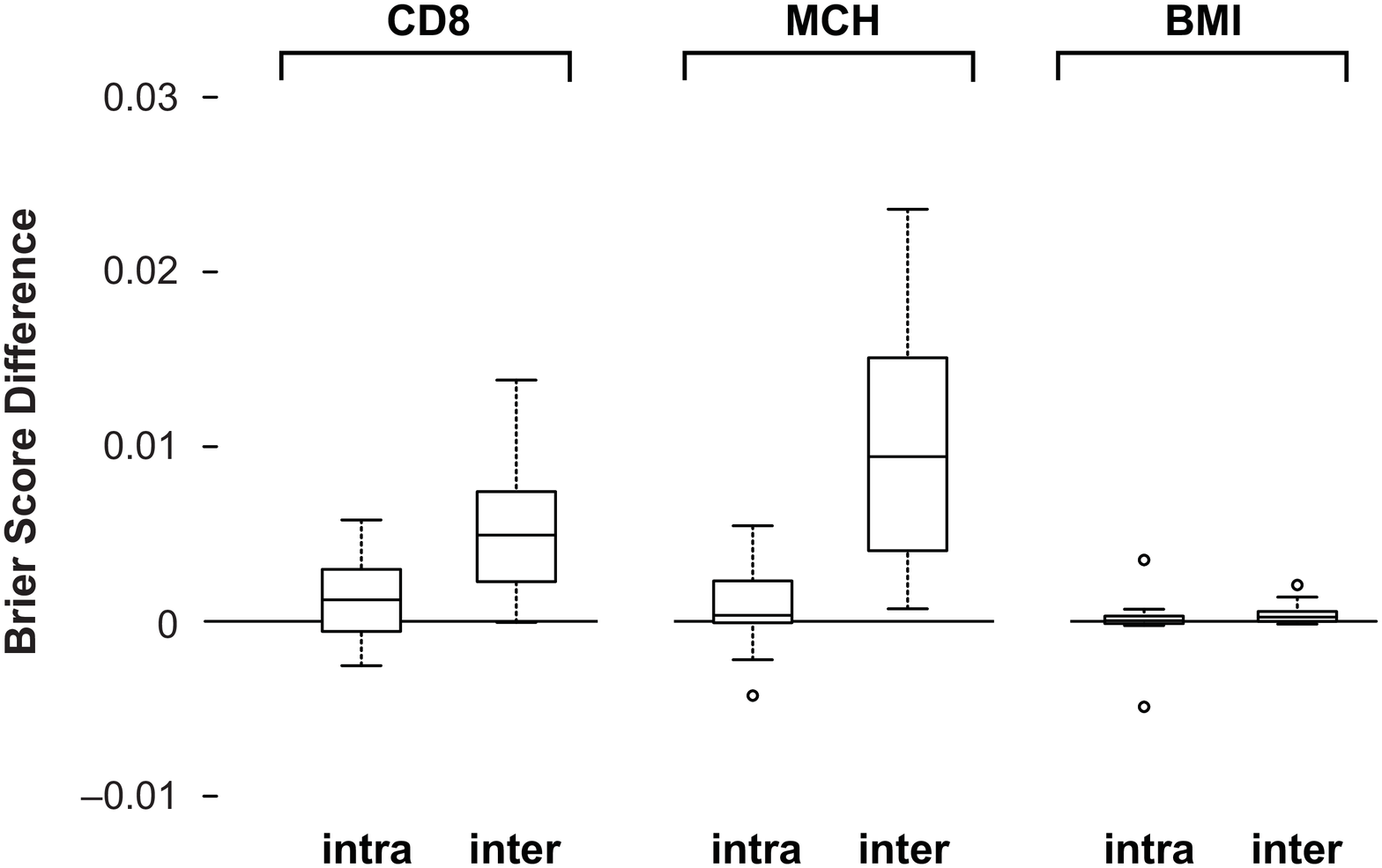}
\caption{Comparison of prediction performance between BSLMM and probit BSLMM for three binary traits in the heterogenous stock mouse data set. Performance is measured by Brier score difference with respect to BSLMM, where a positive value indicates worse performance than BSLMM. The x-axis shows two different ways to split the data into a training set and a test set, each with 20 replicates. The mean Brier scores of BSLMM for the six cases are 0.185, 0.205, 0.201, 0.236, 0.245 and 0.249, respectively. }
\label{fig:real_binary}
\end{figure}

\clearpage
\newpage
\section*{Tables}

\begin{table}[!ht]
 %\begin{tiny}
%\begin{center}
\caption{Summary of some effect size distributions that have been proposed for polygenic modeling. }
 \bigskip
 \begin{small}
  \begin{tabular}{ c|c||c|c }
 \multicolumn{2}{c||}{Effect Size Distribution} & Keywords & Selected References \\ 
Name & Formula & & \\
\hline \hline
t & $\beta_i\sim t(0, \nu, \sigma_a^2)$ & BayesA & \cite{Meuwissen:2001, Verbyla:2009, Verbyla:2010, Hayes:2010} \\ \hline
%[27,32,33,40] \\ \hline
%
 
point-t & $\beta_i\sim \pi t(0, \nu, \sigma_a^2)+(1-\pi)\delta_0$ & BayesB, BayesD, BayesD$\pi$ & \cite{Meuwissen:2001, Verbyla:2009, Verbyla:2010, Hayes:2010, Habier:2011} \\ \hline
%[27,32-34,40]\\ \hline
%
 
 t mixture & $\beta_i\sim \pi t(0, \nu, \sigma_a^2)$ & BayesC & \cite{Verbyla:2009, Verbyla:2010} \\
 %[32,33] \\
 %
&$+(1-\pi) t(0, \nu, 0.01\sigma_a^2)$ && \\ \hline

point-normal & $\beta_i\sim \pi \mbox{N}(0, \sigma_a^2)+(1-\pi)\delta_0$ & BayesC, BayesC$\pi$, BVSR & \cite{Habier:2011, Logsdon:2010, Guan:2011} \\ \hline
%[18,19,34] \\ \hline
% 

double exponential & $\beta_i\sim \mbox{DE}(0, \theta )$  & Bayesian Lasso &  \cite{Park:2008, Campos:2009, Makowsky:2011} \\ \hline 
%[28,39,68] \\ \hline
%

point-normal mixture & $\beta_i\sim \pi_1 \mbox{N}(0, \sigma_a^2)+\pi_2 \mbox{N}(0, 0.1\sigma_a^2)$  & BayesR & \cite{Erbe:2012} \\ 
%[35] \\
%
& $+\pi_3 \mbox{N}(0, 0.01\sigma_a^2)+(1-\pi_1-\pi_2-\pi_3)\delta_0$ & & \\ \hline 

normal & $\beta_i\sim \mbox{N}(0, \sigma_a^2)$ & LMM, BLUP, Ridge Regression & \cite{Hoerl:1970, Whittaker:2000, Yang:2010, Makowsky:2011} \\ \hline
%[22,26,28,48] \\ \hline
%

normal-exponential& $\beta_i\sim \mbox{NEG}(0, \kappa, \theta)$ & NEG & \cite{Hoggart:2008}\\
%[16] \\
%
-gamma&&& \\ \hline

normal mixture & $\beta_i\sim \pi \mbox{N}(0, \sigma_a^2+\sigma_b^2)$  & BSLMM  & Present Work \\
&\qquad $+(1-\pi) \mbox{N}(0, \sigma_b^2)$&&
%\hline
\label{tab:balpha}
\end{tabular}
%\end{center}
\\
\\
\\
The reference list contains only a selection of relevant publications. {\bf Abbreviations:} DE denotes double exponential distribution, NEG denotes normal exponential gamma distribution, and other abbreviations can be found in the main text. In the scaled t-distribution, $\nu$ and $\sigma_a^2$ are the degree of freedom parameter and scale parameter, respectively. In the DE distribution, $\theta$ is the scale parameter. In the NEG distribution, $\kappa$ and $\theta$ are the shape and scale parameters, respectively. {\bf Notes:} {\bf 1.} Some applications of these methods combine a particular effect size distribution with a random effects term, with covariance matrix $K$, to capture sample structure or relatedness. If $K \propto XX^T$ then this is equivalent to adding a normal distribution to the effect size distribution. The listed effect size distributions in this table do not include this additional normal component. {\bf 2.} BayesC has been used to refer to models with different effect size distributions in different papers. {\bf 3.} In some papers, keywords listed here have been used to refer to fitting techniques rather than effect size distributions.  
\end{small}
\end{table}

\clearpage
\newpage

\begin{table}[!ht]
%\begin{center}
\caption{PVE and PGE estimates for five human traits.}
\label{tab:real_pve}
 \bigskip
  \begin{tabular}{ ccccccc }
  & Method & Height & HDL & LDL & TC & TG \\
\hline
\multirow{3}{*}{PVE} & LMM & 0.42 (0.08) & 0.38 (0.15) & 0.22 (0.18) & 0.22 (0.17) & 0.34 (0.17)\\
&BVSR & 0.15 (0.03) & 0.06 (0.01) & 0.10 (0.08) & 0.15 (0.07) & 0.05 (0.06)\\
&BSLMM & 0.41 (0.08) & 0.34 (0.14) & 0.22 (0.14) & 0.26 (0.14) & 0.29 (0.17) \\
\hline
PGE & BSLMM & 0.12 (0.13) & 0.21 (0.14) & 0.27 (0.26) & 0.46 (0.30) & 0.18 (0.20) \\
\end{tabular}
\\
\\
\\
PVE estimates are obtained using LMM, BVSR and BSLMM, while PGE estimates are obtained using BSLMM. Values in parentheses are standard error (for LMM) or standard deviation of posterior samples (for BVSR and BSLMM). $n=3,925, p=294,831$ for height, and $n=1,868, p=555,601$ for other four traits.
%\end{center}
\end{table}

\clearpage
\newpage

\begin{table}[!ht]
%\begin{center}
\caption{Mean computation time, in hours, of various methods for the mouse data set.}
\bigskip
\label{tab:comptime_mouse}
  \begin{tabular}{ c|c }
  Method &Time in hrs (sd) \\ \hline
  LMM & 0.007 (0.002) \\
  LMM-Bayes & 0.14 (0.02) \\
BSLMM-EB & 2.44 (3.52) \\
BSLMM & 3.86 (4.13) \\
 BVSR & 9.20 (6.36) \\
BayesC$\pi$ & 39.4 (11.7) \\
BayesA-Flex & 68.6 (8.06) \\
Bayesian Lasso  & 78.6 (23.4) 
\end{tabular}
%\end{center}
\\
\\
\\
Values in parentheses are standard deviations. Means and standard deviations are calculated based on 2.1 million MCMC iterations in 120 replicates: 20 intra-family and 20 inter-family splits for three phenotypes. Computations were performed on a single core of an Intel Xeon L5420 2.50 GHz CPU. Since computing times for many methods will vary with number of iterations used, and we did not undertake a comprehensive evaluation of how many iterations suffice for each algorithm, these results provide only a very rough guide to the relative computational burden of different methods.
\end{table}

\clearpage
\newpage

 \begin{table}
 \begin{small}
%\begin{center}
\caption{Mean computation time, in hours, of BVSR and BSLMM in all examples used in the present study.}
\label{tab:comptime}
  \begin{tabular}{c|cccccc}
\multirow{2}{*}{Simulations, PVE = 0.2 }
& \multicolumn{6}{c}{Number of Causal SNPs} \\  
  & 10  & 100  & 1000  & 10000  & 10/10000  & 100/10000 \\
\hline
BVSR & 8.25 (1.11) & 50.0 (17.3) & 46.3 (25.0) & 32.8 (25.0) & 50.6 (23.9) & 35.2 (24.7) \\
BSLMM & 8.44 (2.04) & 16.9 (4.06) & 5.44 (0.23) & 5.19 (0.20) & 29.3 (17.2) & 37.5 (19.1)\\
\end{tabular}
\bigskip
\begin{tabular}{c|cccccc}
\multirow{2}{*}{Simulations, PVE = 0.6 }
& \multicolumn{6}{c}{Number of Causal SNPs} \\  
& 10  & 100  & 1000  & 10000  & 10/10000  & 100/10000 \\
\hline
BVSR & 7.43 (0.72) & 36.9 (5.71) & 118 (8.94) & 96.1 (12.0) & 75.4 (28.9) & 92.8 (15.0)\\
BSLMM & 7.34 (1.05) & 41.2 (5.12) & 39.8 (15.0) & 5.13 (0.34) & 13.5 (5.52) & 45.9 (16.6)\\
\end{tabular}
\begin{tabular}{c|ccccccc }
\multirow{2}{*}{Real Data} & \multicolumn{7}{c}{Diseases} \\
& BD  & CAD  & CD  & HT  & RA  & T1D & T2D\\
\hline
BVSR & 131 (20.3) & 110 (20.3) & 105 (13.9) & 114 (21.0) & 12.4 (3.52) & 14.8 (2.47)& 123 (22.0) \\
BSLMM & 21.7 (13.6) & 22.0 (16.2) & 57.6 (24.6) & 32.8 (23.8) & 9.58 (1.17) & 14.5 (2.52) & 38.7 (20.3) \\
\end{tabular}
\bigskip
\begin{tabular}{c|cccccccc}
\multirow{2}{*}{Real Data} & \multicolumn{8}{c}{Quantitative Traits} \\
& Height  & HDL  & LDL  & TC  & TG  & CD8 & MCH& BMI\\
\hline
BVSR & 96.1 & 2.94 & 21.4 & 24.4 & 18.8 & 5.89 (1.87)& 10.5(5.83)& 11.2(8.30)\\
BSLMM & 77.4 & 3.35 & 7.24 & 24.0 & 6.53 & 1.14 (1.05)& 2.45(3.13) & 7.97(3.79)\\
\end{tabular}
%\end{center}
\\
\\
\\
Computations were performed on a single core of an Intel Xeon L5420 2.50 GHz CPU, with 2.1 million MCMC iterations. Values in parentheses are standard deviations. Means and standard deviations are calculated based on 20 replicates in simulations, 20 replicates in the WTCCC data set, and 40 replicates -- 20 intra-family and 20 inter-family splits -- in the mouse data set.
\end{small}
\end{table}

\setcounter{table}{0}
\makeatletter 
\renewcommand{\thetable}{S\@arabic\c@table} 

\clearpage
\newpage

\begin{table}[!ht]
\begin{center}
\caption{Estimates of PVE, PGE and $\log_{10}(\pi)$ for CD8, MCH and BMI in the mouse data set. PVE estimates are obtained using LMM, BVSR and BSLMM, $\log_{10}(\pi)$ estimates are obtained using BVSR and BSLMM, and PGE estimates are obtained using BSLMM. Values in parentheses are standard error (for LMM) or standard deviation of posterior samples (for BVSR and BSLMM). $n=1,410, p=10,768$ for CD8, $n=1,580, p=10,744$ for MCH, and $n=1,828, p=10,771$ for BMI.}
\label{tab:real_pve_mouse}
 \bigskip
  \begin{tabular}{ ccccc }
  & Method & CD8 & MCH & BMI  \\
\hline
\multirow{3}{*}{PVE} & LMM & 0.61 (0.03) & 0.64 (0.03) & 0.14 (0.03)  \\
&BVSR & 0.66 (0.02) & 0.57 (0.02) & 0.13 (0.02) \\
&BSLMM & 0.64 (0.02) & 0.61 (0.03) & 0.13 (0.02)  \\
\hline
PGE & BSLMM & 0.50 (0.09) & 0.43 (0.06) & 0.40 (0.28)  \\
\hline
\multirow{2}{*}{$\log_{10}(\pi)$} & BVSR & -1.69 (0.12) & -1.62 (0.09) & -1.52 (0.21) \\
&BSLMM & -2.68 (0.24) & -2.77 (0.19) & -2.44 (0.70)  \\
\hline
\end{tabular}
\end{center}
\end{table}

\end{document}